\documentclass[a4paper,11pt]{article}

\usepackage{amsfonts,mathrsfs,mathtools,bm,amsmath,xcolor}
\usepackage[margin=25mm]{geometry}
\usepackage{enumerate}
\usepackage{natbib}
\usepackage{booktabs}
\newcommand{\matr}{\mathrm}
\newcommand{\vect}{\mathrm{vec}} 
\newcommand{\gam}{\mathrm{Ga}}

\newcommand{\norm}{\mathrm{N}}
\newcommand{\lognorm}{\mathrm{LN}}
\newcommand{\bet}{\mathrm{Beta}}
\newcommand{\E}{\mathrm{E}}
\newcommand{\invgam}{\mathrm{IG}}

\newcommand{\Var}{\mathrm{Var}}
\newcommand{\Corr}{\mathrm{Corr}}
\newcommand{\diag}{\mathrm{diag}}

\newcommand{\hcauchy}{\mathrm{C}^+}
\newcommand{\tr}{\mathrm{tr}}

\newcommand{\bdiag}{\mathrm{blockdiag}}

\newcommand{\Kappa}{\mathcal{K}}

\renewcommand{\vec}[1]{\boldsymbol{#1}}

\providecommand{\keywords}[1]{\textbf{\textit{Keywords--- }} #1}

\usepackage{xcolor} 

\usepackage{subfig}
\usepackage{graphicx}
\usepackage[export]{adjustbox}
\definecolor{mypurple}{HTML}{5200CC}
\definecolor{myblue}{HTML}{0099FF}
\definecolor{mygrey}{HTML}{A9A9A9}

\usepackage{xr}
\usepackage{authblk}
\title{A sparse Bayesian hierarchical vector autoregressive model for microbial dynamics in a wastewater treatment plant}

\author[1]{Naomi E. Hannaford}
\author[1,*]{Sarah E. Heaps}
\author[1]{Tom M. W. Nye}
\author[2]{Thomas P. Curtis}
\author[1]{Ben Allen}
\author[1]{Andrew Golightly}
\author[1]{Darren J. Wilkinson}
\affil[1]{\small School of Mathematics, Statistics and Physics, Newcastle University, Newcastle upon Tyne, United Kingdom}
\affil[2]{\small School of Engineering, Newcastle University, Newcastle upon Tyne, United Kingdom}
\affil[*]{\small Corresponding author: \texttt{sarah.heaps@ncl.ac.uk}}
\date{}

\begin{document}

\maketitle

\begin{abstract}
Proper function of a wastewater treatment plant (WWTP) relies on maintaining a delicate balance between a multitude of competing microorganisms. Gaining a detailed understanding of the complex network of interactions therein is essential to maximising not only current operational efficiencies, but also for the effective design of new treatment technologies. Metagenomics offers an insight into these dynamic systems through the analysis of the microbial DNA sequences present. Unique taxa are inferred through sequence clustering to form operational taxonomic units (OTUs), with per-taxa abundance estimates obtained from corresponding sequence counts. 
 The data in this study comprise weekly OTU counts from an activated sludge (AS) tank of a WWTP. To model the OTU dynamics, we develop a Bayesian hierarchical vector autoregressive model, which is a linear approximation to the commonly used generalised Lotka-Volterra (gLV) model.  To tackle the high dimensionality and sparsity of the data, they are first clustered into 12 ``bins'' using a seasonal phase-based approach. The autoregressive coefficient matrix is assumed to be sparse, so we explore different shrinkage priors by analysing simulated data sets before selecting the regularised horseshoe prior for the biological application.  We find that ammonia and chemical oxygen demand have a positive relationship with several bins and pH has a positive relationship with one bin. These results are supported by findings in the biological literature. We identify several negative interactions, which suggests OTUs in different bins may be competing for resources and that these relationships are complex. We also identify two positive interactions. Although simpler than a gLV model, our vector autoregression offers valuable insight into the microbial dynamics of the WWTP.
\end{abstract}



\section{Introduction} \label{sec:intro}
Due to recent advances in sequencing technology, there has been an increasing interest in longitudinal studies of microbial communities from a large range of environments.  Unique ecological insights into response to perturbations (or environmental changes) and community stability can be gained from such studies \citep{faust2015}. Furthermore, the complex non-linear interactions between different microbes result in many possible pseudo-stable states \citep{goyal2018}. These interactions and others between different microbes and their environment contribute significantly to microbial dynamics \citep{konopka2015}.

In microbial biology, DNA sequences extracted from environmental samples are grouped together into operational taxonomic units (OTUs) using a clustering algorithm, where typically a single OTU contains sequences that are at least $97\%$ similar with each other \citep{Bunge2014,xia2018}.  Crucially, a particular OTU does not necessarily exactly correspond to a true biological sequence, but OTUs can be thought of as pragmatic proxies for classifying taxa.  We have weekly OTU counts from the activated sludge (AS) of a UK-based wastewater treatment plant (WWTP) over five years, along with corresponding measurements of chemical and environmental (CE) covariates.  After wastewater enters the WWTP, it undergoes the physical process of primary sedimentation, during which large solids are settled out.  The wastewater that emerges from the primary sedimentation tank is called settled sewage and is fed into an aerated tank. The content of the aerated tank is the AS, which plays a pivotal role in wastewater treatment.  Primarily responsible for the consumption of dissolved organic material, it comprises a plethora of different aerobic and anaerobic microorganisms \citep{shchego2016}.  

Microbial communities within AS are complicated biosystems with a network of interconnected trophic links.  For example, to degrade complex polymers, such as proteins, carbohydrates and lipids, many enzymes are required in a multi-stage process.   Several species of microorganisms are needed for complete biodegradation. Gaining theoretical understanding of how these large biological systems work is likely to accelerate the creation of better biotechnological procedures \citep{curtis2003}.  

In the literature, community dynamics are often described by the generalised Lotka-Volterra (gLV) \citep{lotka1926,volterra1926} differential equations, where changes in microbial counts are modelled as a function of taxon-specific growth rates and pairwise interactions.   For example, \citet{mounier2008} used a gLV model to identify interactions within a cheese microbial community.  The gLV model is used to characterise the dynamics of a $K$-species system, where $K>2$. Changes in population of species $i$ are described by
\begin{equation} \label{eq:gLV}
\frac{d}{dt}y_i(t) = b_i y_i(t) + y_i(t)\sum^K_{j=1}a_{ij}y_j(t),
\end{equation}
where $y_i(t)$ is the population size of species $i$ at time $t$, $b_i$ is the growth rate of species $i$ in the absence of any competition and $\matr{A} = \left(a_{ij} \right)$ is a matrix of pairwise interactions.

Often in microbiome studies, the problem of finding the gLV parameters is simplified by using \eqref{eq:gLV} to express $\frac{d}{dt}\log y_i(t)$ as a linear function in the $y_j(t)$, and then discretising the left hand side to view the problem of estimating the gLV parameters as one of least squares  \citep{stein2013,fisher2014,bucci2016}.  Despite widespread use of gLV models, \citet{gibbons2017} investigated microbial dynamics in the human gut with a sparse vector autoregressive (VAR) model.  This model offers the advantage over gLV models of allowing quantification of uncertainty by explicitly modelling error. A key difference between the two approaches is that VAR models assume linear dynamics, whereas gLV models assume non-linear dynamics.    

A VAR model of order one (VAR(1)) can be regarded as a linear approximation to the non-linear numerical solution of a Lotka-Volterra system (see Section~S1 of the Supplementary Materials).  Since linear models tend to be easier to fit, we choose a VAR(1) model for the WWTP data.  A VAR(1) model for $K$ OTUs is given by
\begin{equation}\label{eq:VAR(1)}
\vec{y}_t = \vec{\mu} + \matr{A}\left(\vec{y}_{t-1} - \vec{\mu} \right)  + \vec{\epsilon}_t,
\end{equation}
where $\vec{y}_t$ is a $K$-dimensional vector of counts at time $t$, $\matr{A}$ is a $K \times K$ matrix of autoregressive coefficients and $\vec{\epsilon}_t$ is a vector of $K$ normally distributed errors at time $t$, that is $\vec{\epsilon}_t \sim \norm_K\left(\vec{0},\matr{\Sigma}\right)$.

As explained in Section~\ref{sec:clustering}, we use a seasonal phase-based clustering approach to form 12 ``bins'' of OTUs.  As a result, the bins have a circular time-ordering, which means that a particular bin $k$ typically has its peak abundance one month before bin $k-1$ and one month after bin $k+1$.  Therefore, it is unlikely that the previous abundances for all bins influence the current abundance of any particular bin once abundance in the neighbouring bins is known.  A sparse autoregressive matrix containing many zeroes reflects this idea.  Thus, we investigate different shrinkage priors for this parameter and compare the performance of the priors and their corresponding inferential procedures via a simulation study, before selecting a regularised horseshoe prior.  As a by-product of this approach, we extend the work of \citet{piironen} on prior specification in sparse linear models for a univariate response to consider the problem in a multivariate setting.  This involves careful specification of the hyperprior for the global shrinkage parameter using prior information on sparsity. 

The remainder of the paper is structured as follows. In Section~\ref{sec:eda}, we describe the data and present findings from an exploratory analysis.  Section~\ref{sec:clustering} discusses clustering methods and describes a seasonal phase-based approach.  In Section~\ref{sec:model}, we give a full model specification, including an exploration of shrinkage priors for the matrix of autoregressive coefficients. Section~\ref{sec:app} presents the results of applying our model to the WWTP data. Finally, we summarise our findings in Section~\ref{sec:discuss}.
\section{Exploratory analysis} \label{sec:eda}
\subsection{Data description}\label{ssec:data}
We have weekly counts of 9044 different OTUs measured at $N=257$ time points, starting from 1st June 2011.  A week is missing each year for the Christmas period, which we treat as missing at random.  The dimensions of the data present a significant inferential challenge.  Given that the number of OTUs is much larger than the number of time points, fitting joint models to the counts of all OTUs would be computationally prohibitive. Furthermore, it is unclear whether the number of time points would be sufficiently large to detect interactions and there is the common issue of sparsity in the OTU table, that is, the presence of many zeros.  In our data, approximately $90\%$ of the counts are zero. These could correspond to OTUs that enter the system randomly and die out quickly.  However, it is more likely that the zeros can be attributed to insufficient sampling depth.  These two problems are tackled in Section~\ref{sec:clustering}.

Accompanying the OTU table is a taxonomy table containing the kingdom, phylum, class, order, family and genus of every OTU. However, some of the taxonomic ranks for many of the OTUs are missing.  For example, roughly $54\%$ of OTUs do not have a genus assigned.  Table~S2 in the Supplementary Materials shows the proportions of missing data for each taxonomic rank.  The Ribosomal Database Project (RDP) classifier \citep{Wang2007} was used for classification of each OTU. An OTU was classified as \texttt{NA} if its sequence was not sufficiently similar to entries in the RDP database. 

\begin{table}[h]
\centering
\begin{tabular}{p{140mm} l} \hline
\textbf{Covariate} & \textbf{Unit} \\ \hline
Ammonia, Chloride, COD, DO, Fluoride, MLSS, MLVSS, Nitrate, Nitrite, Phosphate, Sulphate & mg/L  \\
pH  & pH  \\
Temperature & Celsius \\ \hline
\end{tabular}
\caption{Chemical and environmental covariates. \label{tab:chems}}
\end{table}

Finally, we have measurements for 13 different CE covariates, shown in Table~\ref{tab:chems}.  Chemical oxygen demand (COD) measures the amount of oxidisable organic matter dissolved in the sample. Dissolved oxygen (DO) is the concentration of dissolved oxygen.  Mixed liquor suspended solids (MLSS) is the concentration of suspended solids in the tank, determined by filtration and drying at a relatively low temperature. The related mixed liquor volatile suspended solids (MLVSS) records the mass of volatile material lost (evaporated) by heating filtered solids at a higher temperature.
Supplementary Table~S3 shows the proportions of missing data for each covariate.  
We describe how we account for the small amount of missing data in Section~\ref{subsec:missingdata_model}.

In the sections that follow, we discuss an exploratory data analysis of the OTU table and taxonomy table.  We then look at all the data together, with a particular focus on finding possible relationships between some of the CE covariates and the OTUs.

\subsection{OTU tables}\label{ssec:EA_OTUs}
A time series plot (Supplementary Figure~S1) of the total number of OTUs recovered shows an absence of trend.  The total number of OTUs recovered denotes the total number of OTUs (sequences) detected in a sample, at a particular time point.  Since there are 9044 different OTUs, it is not possible to analyse every single OTU, so instead we look at the 12 most abundant OTUs based on median abundance. Figure~S2 in the Supplementary Materials shows time series plots for these top 12 OTUs.  OTUs 8, 15 and 28 clearly demonstrate seasonality with peaks occurring roughly once a year.  OTU 1 also appears to have annual peaks, which are more visible when plotted on the log-scale (see Supplementary Figure S3a). Seasonality with annual peaks is also evident from the stacked bar plot (shown in Supplementary Figure~S4).


\subsection{Taxonomy table} \label{ssec:EA_taxo}

Table S1 in the Supplementary Materials shows the unique numbers of kingdoms, phyla, classes, orders, families and genera.  Supplementary Figure S5 shows the time series plot for the top 12 genera based on median abundance.   Missing genera, which are grouped together in a single ``Unknown'' group, represent a large proportion of the total abundance.
\textit{Rhodobacter} clearly shows seasonality with annual peaks in late February/early March.   \textit{Flavobacterium}, \textit{Ferruginibacter} and \textit{Trichococcus} also seem to display seasonality, although for the latter the seasonality is clearer on the log-scale (shown in Supplementary Figure S3b).   Supplementary Figure~S6 shows that the 12 genera account for about 60\% of the total abundance on average at each time point.

Figure~S7 in the Supplementary Materials shows the time series plot for the 12 most abundant classes based on median abundance.  There are hints of seasonality in some of the classes, even at this fairly coarse taxonomic rank, for example, \textit{Flavobacteriia}, \textit{Actinobacteria} and \textit{Deltaproteobacteria} show rough annual peaks.  \textit{Clostridia}, \textit{Bacilli} and \textit{Gammaproteobacteria} also show seasonal behaviour. 
The most abundant class is \textit{Alphaproteobacteria} and its time series profile is very noisy without any obvious annual peaks.  \textit{Alphaproteobacteria} form one of the most abundant groups of bacteria on the planet and are extremely diverse \citep{williams2007}, so it is unsurprising that this class is the most abundant.  The diversity of \textit{Alphaproteobacteria} is also reflected here, as there are 1238 different OTUs from the class present.   A class as diverse as this may have species that prefer different conditions and hence have peaks in population size at different times of the year.  The stacked bar plot for the top 12 classes (Supplementary Figure~S8) shows the dominance of \textit{Alphaproteobacteria} and \textit{Betaproteobacteria}.  

\subsection{Analysis of the combined data}\label{ssec:EA_link}
\begin{figure}[h]
\begin{center}
\subfloat[][]{\label{fig:heatmap_OTUs}\includegraphics[width=0.45\textwidth , valign=b]{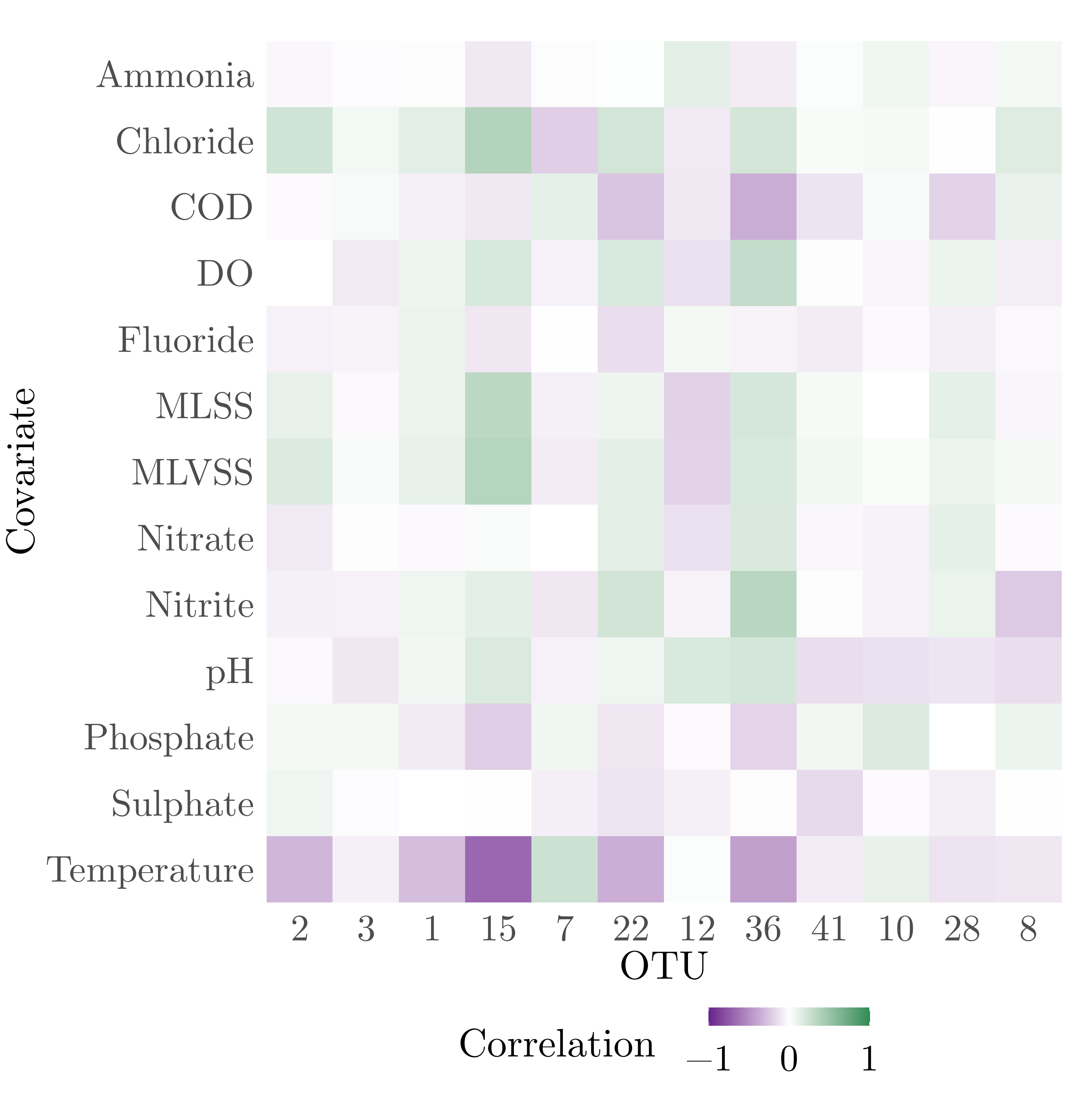}}\hspace{1cm}
\subfloat[][]{\label{fig:heatmap_chems}\includegraphics[width=0.435\textwidth , valign=b]{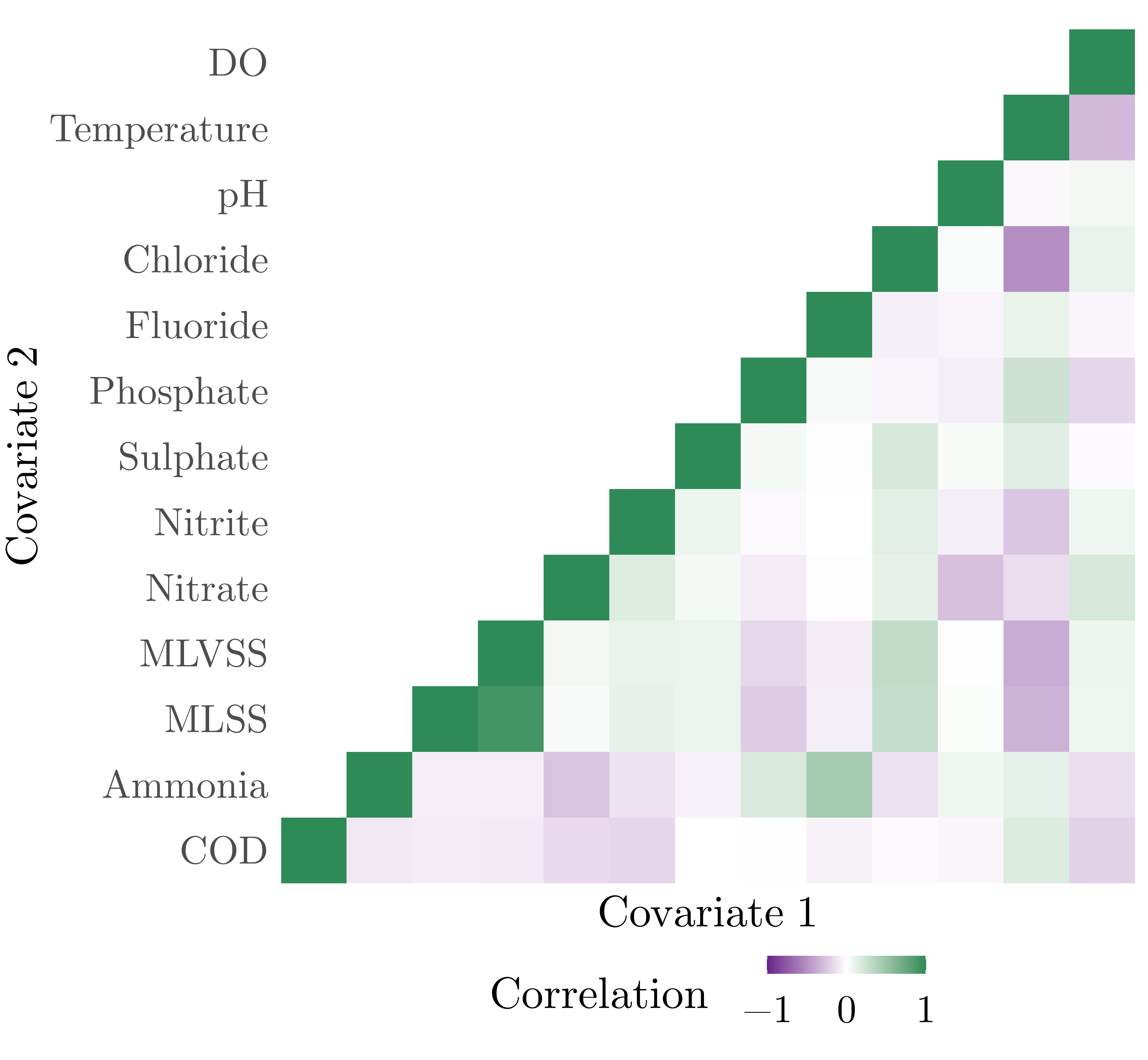}}
\caption{Heatmaps of the pairwise correlations \protect\subref{fig:heatmap_OTUs} between the top 12 OTUs and the CE covariates and \protect\subref{fig:heatmap_chems} among the chemical and environmental covariates.}
\label{fig:heatmaps_otus_chems}
\end{center}
\end{figure}

In this section we investigate relationships between OTU abundance and the CE covariates. We identify potential relationships between the CE covariates and the 12 most abundant OTUs, genera and classes.  Figure~\ref{fig:heatmap_OTUs} shows a heatmap of the correlations between the covariates and the relative abundances of the top 12 OTUs.  Some of the most abundant OTUs appear to be correlated with temperature.   There appear to be some correlations between some of the OTUs and chloride, nitrite, COD, DO, phosphate, MLSS and MLVSS. Figure~\ref{fig:heatmap_chems} shows a heatmap of the pairwise correlations between the CE covariates.  It seems that these covariates are potentially correlated with temperature. However, with the exception of temperature and chloride, these correlations are fairly weak, so it would be na\"{i}ve to attribute all of the possible relationships that we see here to an indirect relationship with temperature.   Microorganisms in AS are very diverse and interact with, feed on and utilise chemical compounds in many different ways.  Nevertheless, it does seem that in general for these top 12 OTUs, if the correlation with temperature is weaker, then the correlations with other covariates tend to be weaker too. 

\begin{figure}[h]
\centering
\subfloat[][]{\label{fig:heatmap_genera}\includegraphics[width=0.45\textwidth , valign=b]{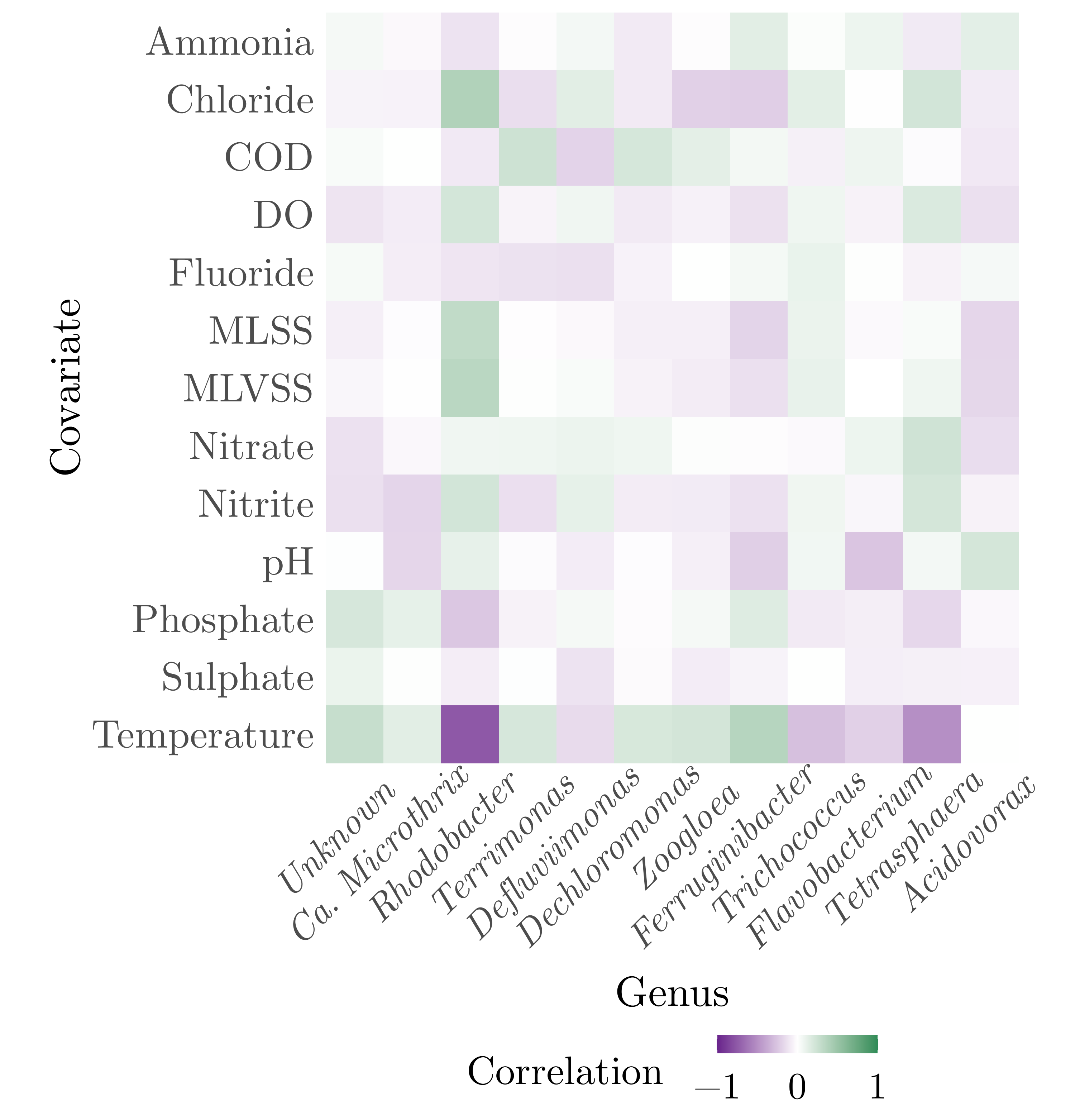}}\hspace{1cm}
\subfloat[][]{\label{fig:heatmap_class}\includegraphics[width=0.43\textwidth , valign=b]{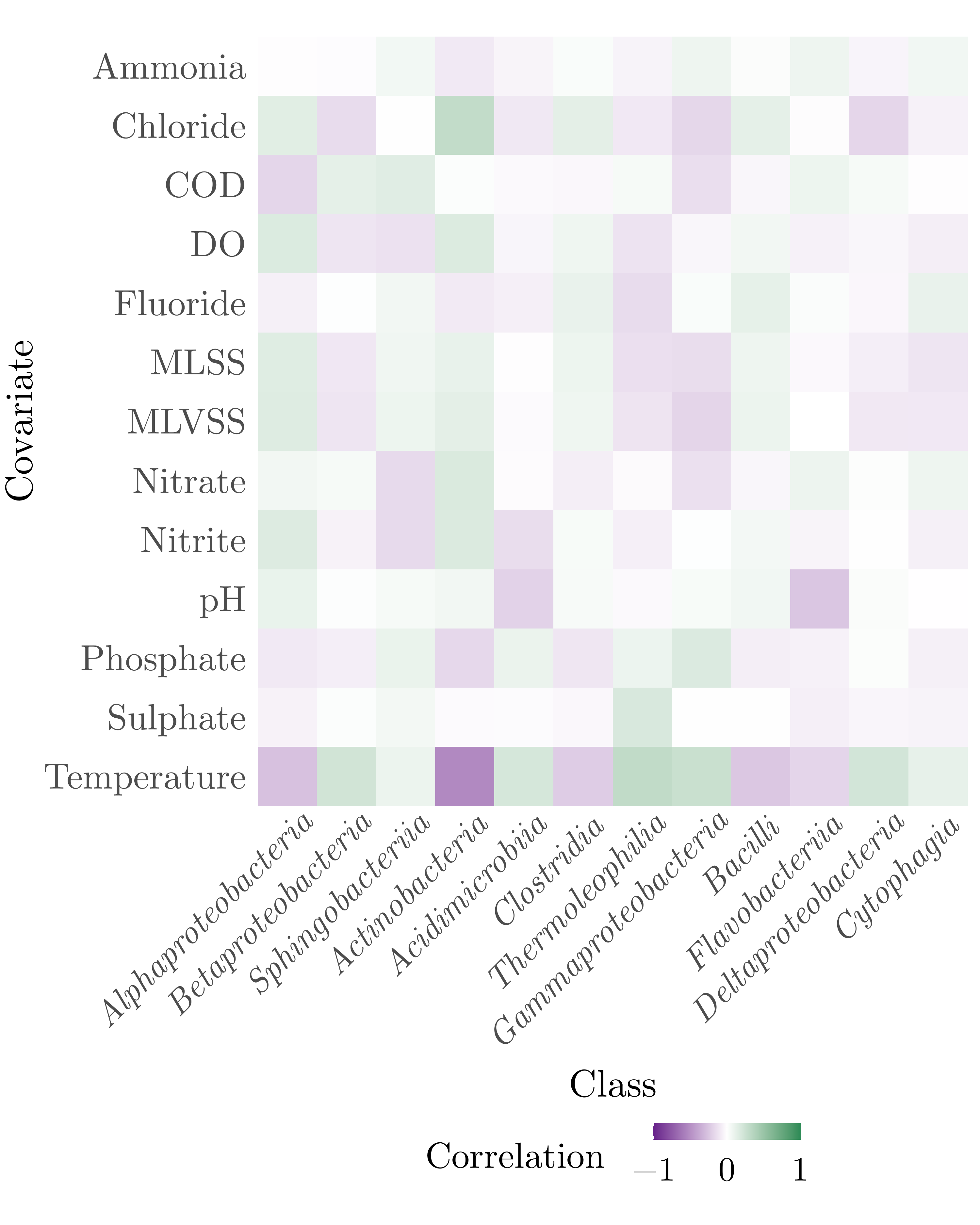}}
\caption{Heatmap of the correlations between the chemical and environmental covariates and \protect\subref{fig:heatmap_genera} the 12 most abundant genera and \protect\subref{fig:heatmap_class} the 12 most abundant classes.}
\label{fig:heatmaps_taxo}
\end{figure}
Figure~\ref{fig:heatmap_genera} shows the heatmap of correlations between the covariates and the 12 most abundant genera.  Most of the genera have at least a weak correlation with temperature, with the exception of \textit{Acidovorax}, \textit{Ca. Microthrix} and \textit{Defluviimonas}.  \textit{Rhodobacter}, \textit{Ferruginibacter} and \textit{Tetrasphaera} appear to have a strong correlation with temperature, which we already identified as showing seasonal behaviour in Section~\ref{ssec:EA_taxo}.  We also see that pH has a weak correlation with \textit{Acidovorax}, \textit{Ferruginibacter} and \textit{Flavobacterium}. There are weak correlations present between: nitrite and \textit{Tetrasphaera}; nitrite and \textit{Rhodobacter}; and nitrate and \textit{Tetrasphaera}. COD and chloride have weak correlations with a few of the genera.

A heatmap of the correlations between the top 12 classes and the CE covariates are shown in Figure~\ref{fig:heatmap_class}.  Most classes exhibit at least weak correlation with temperature.  \textit{Flavobacteriia} appears to have a negative correlation with pH and \textit{Acidimicrobiia} appears to have a weak negative correlation with pH too. 

\subsection{Summary}\label{ssec:EDA_summary}
From this exploratory analysis, we have identified that there are relationships between some of the CE covariates and the relative abundances of some of the top 12 OTUs.  Relationships can also be seen at the coarser taxonomic ranks of genera and classes.  We have also seen that some of the CE covariates are correlated with each other.  Finally, we have observed signs of seasonality and absence of time trend in the relative abundances at both fine and coarse taxonomic ranks. These observations will help to inform decisions when developing our model in Section~\ref{sec:model} and they will also aid interpretation of our results in Section~\ref{sec:app}.


\section{Clustering} \label{sec:clustering}
The data are counts of OTUs, where some OTUs have counts in the thousands and others have (mostly) counts of zero throughout time.  Zeros can arise for structural reasons (``hard zeros'') or due to lack of sampling depth (``soft zeros'') \citep{kaul2017}.  As such, we would expect an excess of zeros over Poisson variation.  Indeed, $91.5\%$ of the counts are zero. Therefore, a natural approach might be to use a time series model for zero-inflated multivariate count data, for example, see \citet{lee2018}. However, there are over 9000 OTUs, indicating our model would have to allow over 81 million pairwise interactions.  To make model-fitting more manageable we instead choose to cluster the data, following analyses by other authors (see \citet{eiler2012,stein2013,david2014,dam}).  Choosing a small enough number of clusters removes the complication of zero inflation and allows us to make the simplifying assumption that our data can be modelled as continuous. 

An approach that reduces the dimensionality of the data, but also retains all of the OTUs, is taxonomy-based clustering.  This involves taking the $n$ most abundant taxa at each time point that represent a high percentage, say $90\%$, of the total abundance and grouping the remaining taxa into an ``other'' category. The term taxa here could refer to any taxonomic rank.  For example, \citet{stein2013} grouped OTUs into the top ten genera and an ``other'' category in their work to infer gut microbiota ecology in mice. 
This approach is unsuitable for our data because of the large proportion ($54.1\%$) of missing taxonomic information at the genus level and the large number of genera ($187$) required to capture $90\%$ of the abundance.  Even when considering median abundance and coarser taxonomic ranks, we find that the finest taxonomic rank we can use without having an unknown as a group is class, which is possibly too coarse. As we noted in Section~\ref{ssec:EA_taxo}, \textit{Alphaproteobacteria} was the most abundant class in the AS tank (with 1238 different OTUs) but this class is known to be extremely diverse in general \citep{williams2007}.  Modelling the change in its abundance over time and its interactions with other classes and the environment is unlikely to yield biologically useful insight, given that the different OTUs within the class may prefer different conditions. 

\citet{dam} researched dynamic models of the complex microbial metapopulation in a lake and suggested that, for characterising interaction dynamics, clustering by taxonomy is not an effective strategy. 
 They proposed an alternative method of clustering OTUs, where they define peak profiles, which involves identifying positions in time where each OTU has its largest abundance(s). The OTUs are then clustered into ``subcommunities'' based on these profiles with remaining OTUs placed in an additional group. The rationale is that these subcommunities represent OTUs with similar dynamics perhaps because of symbiotic relationships or shared dependence on the environment.  Since we have clear evidence of seasonality in the WWTP data, we adopt a similar approach. 

  \subsection{Time series clustering\label{ssec:ts_clustering} }
  \begin{figure*}[htb]
\centering
\includegraphics[width=0.75\textwidth]{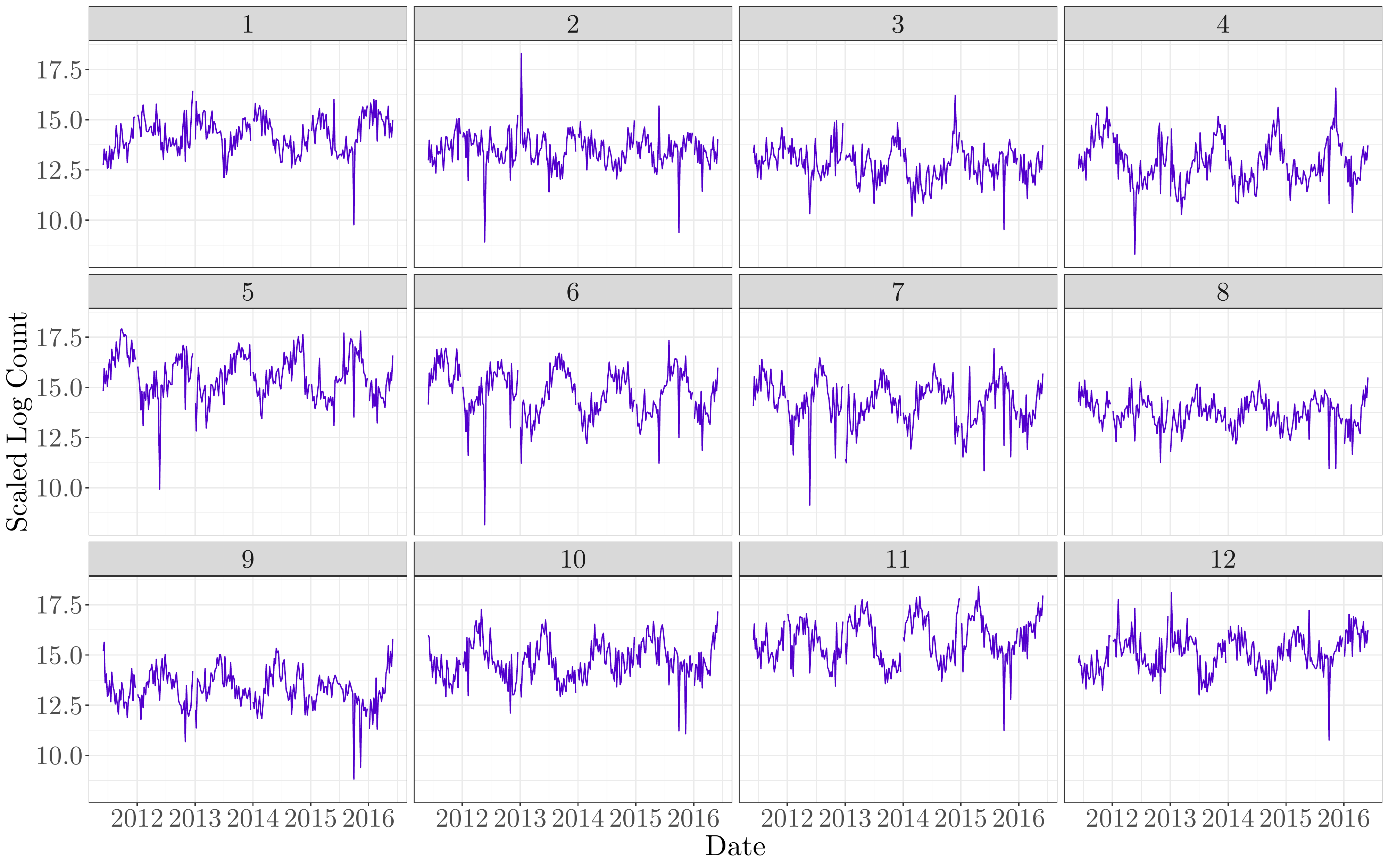}
\caption{Time series plots of the scaled log counts for the 12 bins.}
\label{fig:AS_bins}
\end{figure*}
We adopt the time series clustering method as follows.  First, we calculate the scaled weekly means of each OTU.  Then we represent the annual series for each OTU using a Fourier basis and calculate the phase and amplitude of each harmonic, where the frequency is $2\pi kt/51$  for the $k$-th harmonic and time $t$.  We denote the first phase of OTU $i$ by $\phi_i$.  The interval $\left[-\pi, \pi \right]$ is divided into $12$ equally sized intervals and we assign OTU $i$ to the interval in which $\phi_i$ lies for all $i$. This gives 12 clusters, which we call ``bins''.  Let $\tilde{w}_{ti}$ be the count of OTU $i$ at time point $t$.  The set of OTUs in bin $j$ is $S_{j}$ and $w_{tj} = \sum_{i \in S_j} \tilde{w}_{ti} $ is the count for bin $j$ at time $t$. Visual inspection of the counts of each bin reveals that the bins peak once per year, with different bins peaking in different months.  

To stabilise the variance of the $w_{tj}$ series over the year we log-transform the counts of the bins and set $\tilde{y}_{tj} = \log(x_{tj}).$  We then scale the log counts of each bin so that their variance is roughly one (see Section~\ref{ssec:meff}), denoting by $y_{tj} = \tilde{y}_{tj}/\bar{s}$ for all $t$ and $j = 1,\dots,12$, where $\bar{s} = (\sum^{12}_{j=1} s_j)/12$, $s_j$ is the standard deviation of $\tilde{y}_{1:N,j}$ and $y_{tj}$ denotes the scaled log counts.   Figure~\ref{fig:AS_bins} shows the time series plots of the scaled log counts for the 12 bins.  The plots do not seem to demonstrate any signs of a time trend, which could indicate a mean net growth rate of zero.  Each bin clearly shows seasonal behaviour with a peak every year, with the exception of bins 2 and 3, where the peaks are not as obvious. We can see that for each bin the annual peaks are different.  Bin 1 seems to peak in February, bin 2 seems to peak in January, bin 3 seems to peak in December and so on.  This labelling of the bins is a consequence of the first time point in the series being at the start of the June and the interpretation of the phase for each bin.  For $\phi=0$ we have a harmonic (sine wave) that is $0$ at time $t=0$, which roughly corresponds to the last week of May.  The peak of this harmonic will be at the end of August. Now, for example, take bin 1, which contains all OTUs with $\phi \in \left[-\pi,-5\pi/6\right)$.  This corresponds to the harmonic being shifted five to six months forward in time (to the right) and means that OTUs in this bin typically peak anywhere between the end of January and February.  

\section{Model description}\label{sec:model}

As discussed in Section~\ref{sec:intro}, a vector autoregression is chosen as the model for the clustered OTU data. The novelty in our approach lies in the use of a hierarchical prior for the autoregressive coefficient matrix and error variance matrix. This is constructed to allow sparsity in the former whilst facilitating the development of a principled methodology for the specification of the hyperparameters in the prior.

It is sometimes proposed that microbes in the AS of WWTPs are in a stable state, after allowing for environmental changes over time \citep{shchego2016}.  Imposing stability of the dynamic system as a model assumption could be achieved by constraining inference for the autoregressive coefficient matrix $\matr{A}$ to the stationary region, that is, the region where the spectral radius of $\matr{A}$ is less than one. However, constructing a sparsity-inducing prior that respects the geometry of the stationary region is non-trivial \citep[][]{heaps20}. Therefore we leave $\matr{A}$ unconstrained but observe in our later analysis in Section~\ref{sec:app} that all the posterior samples lies within the stationary region.  

\subsection{Error structure}\label{ssec:error}
When modelling with a vector autoregression, it is common to adopt a parsimonious parametric form for the error variance matrix $\matr{\Sigma}$, for example, by assuming that $\matr{\Sigma}$ is diagonal, that is $\matr{\Sigma} = \sigma^2 \matr{I}_K$, where $\matr{I}_K$ represent a $K \times K$ identity matrix. Since the clustered OTU data have a circular time-ordering, with the OTUs in neighbouring bins (or bins 12 and 1) typically peaking in abundance in neighbouring months, a more appropriate parametric form is to assume $\matr{\Sigma}^{-1}$ is a symmetric, circulant, tridiagonal matrix of the form
\begin{equation} \label{eq:tri_diag_matrix}
\matr{\Sigma}^{-1} =
\begin{pmatrix}
\sigma_0^{-2} &\omega &0        &0        &\cdots &0      &0        &0        &\omega\\
\omega &\sigma_0^{-2} &\omega &0        &\cdots &0      &0        &0        &0\\
0        &\omega &\sigma_0^{-2} &\omega &\cdots &0      &0        &0        &0\\
\vdots   &\vdots   &\vdots   &\vdots   &\ddots &\vdots &\vdots   &\vdots   &\vdots\\
0        &0        &0        &0        &\cdots &0      &\omega &\sigma_0^{-2} &\omega\\
\omega &0        &0        &0        &\cdots &0      &0        &\omega &\sigma_0^{-2}
\end{pmatrix}.
\end{equation}
This is the precision structure of a circular first-order autoregressive model \citep[][Chapter 13]{gelfand2010} for which the interpretation of the errors, at any time $t \in \{1,\dots,N\}$ is as follows: (i) for $j < k$, the errors of $y_{tk}$ have stronger correlations with the errors of $y_{tj}$ when $\min\{ k-j, j+K-k \}$ is smaller; (ii) the correlation between the errors of $y_{tj}$ and $y_{t,j+\ell}$ is the same as the correlation between the errors of $y_{tj}$ and $y_{t,j-\ell}$, where $j+\ell$ and $j-\ell$ are in arithmetic modulo $K$ (and $0$ is written as $K$ rather than $0$). 

To ensure that $\matr{\Sigma}^{-1}$  is  positive definite, it is convenient to reparameterise in terms of $\sigma_0^{-2} = \left( \varpi_0 + \varpi_1 \right)/\sqrt{2}$ and $\omega = \left(\varpi_0 - \varpi_1 \right)/2 \sqrt{2}$. The precision matrix is then positive definite if and only if $\varpi_i > 0$ for $i=0,1$. The full derivation of this is given in Section~S3 of the Supplementary Materials.

\subsection{Time varying mean}
Due to the manner in which the OTU data were clustered, it is likely that a time varying mean will be more appropriate than a mean which is static over time. This was also evident from the time series plots of the bins in Figure~\ref{fig:AS_bins}. We therefore modify the simple VAR(1) model in~\eqref{eq:VAR(1)} to give
\begin{equation}\label{eq:VAR(1)_with_mean}
\vec{y}_t = \vec{\mu}_t + \matr{A}(\vec{y}_{t-1} -\vec{\mu}_{t-1}) + \vec{\epsilon}_t, 
\end{equation}
for $t=2,\ldots,N$, where $\matr{A}$ is unstructured and $\vec{\epsilon}_t$ is normally distributed with zero-mean and a precision matrix of the form \eqref{eq:tri_diag_matrix}. To capture the seasonal variation of each bin, we use a harmonic regression to fit a time varying mean, that is
\begin{equation}
\vec{\mu}_t = \vec{\alpha}_t + \sum^{J}_{j=1} \vec{\beta}_j \sin\left(\frac{2\pi tj}{52}\right) + \vec{\gamma}_j \cos\left(\frac{2\pi tj}{52}\right), \label{eq:mean}
\end{equation}
where $J$ is the number of harmonics. After fitting the model with $J = 1,\dots,4$ harmonics, we select $J=2$ for our final model because there was little evidence of $\vec{\beta}_j$ and $\vec{\gamma}_j$  being non-zero for $j=3,4$.   We then capture the information from the CE covariates by expressing the intercept term at time $t$ $\vec{\alpha}_t$ as a linear combination of the CE covariates at the previous time point $t-1$. This was based on expert biological judgement, motivated by the idea that the effect of any environmental conditions is unlikely to be instantaneous.  

Let $\tilde{\matr{X}}$ be a $N \times L$ matrix of covariates.  We find the covariate data are skewed and so apply a square-root transformation.  This transformation relates to how we handle missing data (see Section~\ref{subsec:missingdata_model}). Each column (covariate) is then standardised, denoting the resulting matrix as $\matr{X}$. We let
\begin{equation}
\alpha_{tk} = b_{0k} + b_{1k}x_{t-1,1} + \dots + b_{Lk}x_{t-1,L}, \label{eq:alpha}
\end{equation}
where $x_{t\ell}$ is the measurement of covariate $\ell$ at time $t$ and $b_{\ell k}$ is a regression coefficient for bin $k$ and covariate $\ell$, noting that $b_{0k}$ is the intercept term for bin $k$. We collect the $b_{ij}$ into a $(L+1) \times K$ matrix $\matr{B}$ in which $b_{ij}$ is the $(i+1,j)$-element.

To select which covariates to include, we fit the model without any CE covariates so that $\alpha_{tk}=b_{0k}$ for $t=2,\ldots,N$ and $k=1,\ldots,K$.  Then we check to see which covariates have a lag-one correlation with the posterior mean of the model residuals at each time point. This is a convenient yet simple method for variable selection. Though a more rigorous approach might instead make use of a shrinkage prior, our focus in this paper is sparsity in the matrix of autoregressive coefficients, as we discuss in Section~\ref{sec:sparsity}. We find that $L=5$ covariates are correlated with the residuals: nitrate, chemical oxygen demand (COD), ammonia, pH and phosphate.  This selection of five covariates is supported by our exploratory data analysis.  In Section~\ref{ssec:EA_link}, we found that several of the top 12 OTUs appeared to have a (contemporary) correlation with COD, ammonia and phosphate. We also found that one of the top 12 genera seemed to be correlated with nitrate and two of the top 12 classes had weak negative (contemporary) correlations with pH.

\subsection{Missing data model}\label{subsec:missingdata_model}

As discussed in Section~\ref{ssec:data}, there are missing data for some of the CE covariates. As with the missing values of $y_{tj}$, which are treated as missing at random, we treat these missing values as unknowns and average over our uncertainty in their values. This requires specification of a model for the (transformed) covariates $\matr{X}$.  We define $\vec{x}_t = (x_{t1},\hdots,x_{tL})^T$ and allow the covariates to evolve according to a simple first-order autoregression
\begin{equation}\label{eq:missingdata_model}
\vec{x}_t = \matr{\Phi}_{X}{\vec{x}}_{t-1} + \tilde{\vec{\epsilon}}_t, \quad \tilde{\vec{\epsilon}}_t \sim \norm_L\left(\vec{0}, \matr{\Sigma}_{X}\right),\\
\end{equation} 
for $t = 2, \hdots, N$, where $\matr{\Phi}_X$ is assumed to be diagonal, $\matr{\Phi}_X = \diag(\phi_{X,1},\ldots,\phi_{X,L})$.

\section{Sparsity in the autoregressive coefficient matrix}\label{sec:sparsity}

The clustered OTU data comprise $N=257$ observations on $K=12$ bins and yet there are $K^2=144$ unknown coefficients in the autoregressive coefficient matrix in~\eqref{eq:VAR(1)_with_mean}. Given the circular time-ordering of the bins in the clustered OTU data, the previous abundances for all bins are unlikely to influence the current abundance of any particular bin once the abundances of neighbouring bins are known. A sparse autoregressive matrix would reflect this notion, allowing the more influential coefficients to be learnt with a greater degree of precision, whilst providing useful biological interpretation of the sparse matrix structure. Since the parameters in the autoregressive coefficient matrix can be interpreted as regression coefficients in a linear model, this is essentially a problem of variable selection which can be addressed by assigning a sparsity-inducing prior to the elements in the autoregressive coefficient matrix; see, for example, \citet{Gef14} or \citet{ABC16}. A commonly adopted prior is a zero-mean scale-mixture of normals, in which the mixing distribution can either be discrete, as in spike-and-slab priors \citep[][]{MB88,GM93}, or continuous, as in the (regularised) horseshoe \citep[][]{carvalho09,carvalho10,piironen}. A simulation experiment, described in Section~S4 of the Supplementary Materials, suggests that the horseshoe prior can yield very similar inferences to the spike-and-slab prior, but with greatly improved mixing during posterior sampling, and so we adopt a prior of this form for the parameters of the autoregressive coefficient matrix. 

The horseshoe prior belongs to a class of global-local shrinkage priors in which the effect of a global hyperparameter is to encourage shrinkage of all coefficients towards zero. Local shrinkage hyperparameters, in one-to-one correspondence with the coefficients, then give the prior a heavy tail so as to retain support for large values of individual coefficients. It is widely known that the posterior can be very sensitive to the prior chosen for the global shrinkage parameter. Motivated by this observation, \citet{piironen} describe a principled methodology for the specification of this prior in the context of (generalised) linear regression. However, their attention was limited to univariate response variables. In order to apply similar ideas to the important choice of prior for the global shrinkage parameter in the vector autoregressive model, the remainder of this section describes an extension of the methodology to the class of linear regression models with a multivariate response, of which a vector autoregression can be regarded as a special case.

\subsection{The shrinkage factor matrix}\label{subsec:shrinkage_factor}

Denote by $\vec{y}_i = (y_{i1},\ldots,y_{iQ})^T$ a $Q$-variate response vector and by $\vec{x}_i$ a $P$-variate vector of covariates for experimental unit $i$. Under the multivariate linear regression model
\begin{equation}\label{eq:multi_reg}
\vec{y}_i = \matr{A}^T \vec{x}_i + \vec{\epsilon}_i, \quad \vec{\epsilon}_i \sim \norm_Q(\vec{0}, \matr{\Sigma}), \quad i=1,\ldots,N,
\end{equation}
where $\matr{A} = (a_{jk})$ is a $P \times Q$ matrix of regression coefficients and $\vec{\epsilon}_i = (\epsilon_{i1},\ldots,\epsilon_{iQ})^T$ is a vector of errors. We recover the order 1 vector autoregressive model when $i$ indexes time, $\vec{x}_i=\vec{y}_{i-1}$ and $Q=P=K$. In matrix form,~\eqref{eq:multi_reg} can be written as
\begin{equation}\label{eq:multi_reg_mat}
\matr{Y} = \matr{X} \matr{A} + \matr{E}
\end{equation}
where $\matr{Y}$ and $\matr{X}$ are $N \times Q$ and $N \times P$ data matrices with $i$-th rows $\vec{y}_i^T$ and $\vec{x}_i^T$, and $\matr{E}$ is a $N \times Q$ matrix of errors with $i$-th row $\vec{\epsilon}_i^T$.

Stacking the rows of $\matr{Y}$ and $\matr{A}$ into $NQ$- and $PQ$-vectors, respectively, let $\vec{y}^* = \vect( \matr{Y}^T ) = (y_{11}, \ldots, y_{1Q}, y_{21}, \ldots, y_{NQ})^T$ and $\vec{a}^* = \vect( \matr{A}^T ) = (a_{11}, \ldots, a_{1Q}, a_{21}, \ldots, a_{PQ})^T$. Now assume we give the regression coefficients $\vec{a}^*$ a horseshoe prior
\begin{align}
\begin{split}
a_{jk} | \lambda_{jk}, \tau &\sim \norm(0, \tau^2 \lambda_{jk}^2),\label{eq:horseshoe}\\
\lambda_{jk} &\sim \hcauchy(0, 1), \quad j=1,\ldots,P, \; k=1,\ldots,Q,
\end{split}
\end{align}
in which the $\lambda_{jk}$ are the local shrinkage parameters, $\tau$ is the global shrinkage parameter and $\hcauchy(a,b)$ denotes a half Cauchy distribution with location $a$ and scale $b$. 

In Section~S5.1 of the Supplementary Materials, we show that the conditional posterior for the regression coefficients $\vec{a}^*$ given the shrinkage parameters, $\matr{\Lambda}^* = \diag(\lambda_{11}, \ldots, \lambda_{1Q}, \lambda_{21}, \ldots, \lambda_{PQ})^T$ and $\tau$, along with the error variance $\matr{\Sigma}$, is given by
\begin{equation}\label{eq:cond_posterior_a*}
\vec{a}^* | \matr{\Lambda}^*, \tau, \matr{\Sigma}, \vec{y}^* \sim \norm_{PQ} (\vec{m}^*, \matr{V}^*)
\end{equation}
where  
\begin{equation*}
\vec{m}^* = \tau^2 \matr{\Lambda}^* \left[ \tau^2 \matr{\Lambda}^* + \left\{ \left( \matr{X}^T \matr{X} \right)^{-1} \otimes \matr{\Sigma} \right\} \right]^{-1} \hat{\vec{a}}^*, \qquad \matr{V}^*  = \left\{ \tau^{-2} \matr{\Lambda}^{*\, -1} + \left( \matr{X}^T \matr{X} \otimes \matr{\Sigma}^{-1} \right) \right\}^{-1}
\end{equation*}
in which
\begin{equation*}
\hat{\vec{a}}^* = \vect(\hat{\matr{A}}^T) = \left\{ \left( \matr{X}^T \matr{X} \right)^{-1} \matr{X}^T \otimes \matr{I}_Q \right\} \vec{y}^*
\end{equation*}
and $\hat{\matr{A}}$ is the least squares estimator of $\matr{A}$.

If we assume that the explanatory variables are uncorrelated with zero mean and variance $\Var(X_j) = s_j^2$, then $\matr{X}^T \matr{X} \simeq N \diag(s_1^2, \ldots, s_P^2)$ and so
\begin{align*}
\tau^2 \matr{\Lambda}^* \left[ \tau^2 \matr{\Lambda}^* + \left\{ \left( \matr{X}^T \matr{X} \right)^{-1} \otimes \matr{\Sigma} \right\} \right]^{-1}  &= \bdiag\left\{ \tau^2 \matr{\Lambda}_1 \left( \tau^2 \matr{\Lambda}_1 + \frac{1}{N s_1^2} \matr{\Sigma} \right)^{-1}, \ldots, \tau^2 \matr{\Lambda}_P \left( \tau^2 \matr{\Lambda}_P + \frac{1}{N s_P^2} \matr{\Sigma} \right)^{-1} \right\}\\
&= \bdiag\left( \matr{I}_Q - \Kappa_1, \ldots, \matr{I}_Q - \Kappa_P \right)
\end{align*}
where $\matr{\Lambda}_j = \diag(\lambda_{j1}^2, \ldots, \lambda_{jQ}^2)$ is the $j$-th diagonal block of $\matr{\Lambda}^*$ and
\begin{equation*}
\Kappa_j = \left( \matr{I}_Q + N s_j^2 \tau^2 \matr{\Lambda}_j \matr{\Sigma}^{-1} \right)^{-1}, \quad j = 1,\ldots,P.
\end{equation*}
Hence we have
\begin{equation*}
\vec{m}^* = \bdiag\left( \matr{I}_Q - \Kappa_1, \ldots, \matr{I}_Q - \Kappa_P \right) \hat{\vec{a}}^* = \vect(\matr{M}^T)
\end{equation*}
where $\matr{M} = (m_{jk}) = \E(\matr{A} | \matr{\Lambda}^*, \tau, \matr{\Sigma}, \matr{Y})$. If we define $\vec{a}_j = (a_{j1}, \ldots, a_{jQ})^T$, $\hat{\vec{a}}_j = (\hat{a}_{j1}, \ldots, \hat{a}_{jQ})^T$ and $\vec{m}_j = (m_{j1}, \ldots, m_{jQ})^T$ for $j=1,\ldots,P$ as the (transposed) columns of $\matr{A}$, $\hat{\matr{A}}$ and $\matr{M}$, respectively, then it is clear that
\begin{equation*}
\vec{m}_j = \left( \matr{I}_Q - \Kappa_j \right) \hat{\vec{a}}_j, \quad j=1,\ldots,P
\end{equation*}
and so we can imagine constructing the posterior mean $\matr{M}^T$ of $\matr{A}^T$ column-wise; column $j$, corresponding to the coefficients of covariate $j$ in the linear predictors of $Y_1$ through $Y_Q$, is a linear transformation of column $j$ of the (transposed) least squares estimator $\hat{\matr{A}}^T$.

Since $\matr{\Lambda}_j$ and $\matr{\Sigma}$ are real and positive definite, the eigenvalues, $\eta_1, \ldots, \eta_Q$, of $N s_j^2 \tau^2 \matr{\Lambda}_j \matr{\Sigma}^{-1}$ must be real and positive. The eigenvalues of $\Kappa_j^{-1} = \matr{I}_q + N s_j^2 \tau^2 \matr{\Lambda}_j \matr{\Sigma}^{-1}$ are therefore $1 + \eta_j > 1$ for $j=1,\ldots,Q$, and hence the eigenvalues of $\Kappa_j$, $1/(1+\eta_j)$, must lie between 0 and 1, making it a \emph{convergent matrix}. We can therefore regard $\Kappa_j$, as the \emph{shrinkage factor matrix} for coefficients $\vec{a}_j$ of covariate $j$. The size of the eigenvalues of $\Kappa_j$ determine the extent to which the coefficients $\vec{a}_j$ are shrunk towards zero. Since the eigenvalues $\eta_j$ are directly proportional to $\tau^2$, as $\tau \to 0$, all eigenvalues of $\Kappa_j$ approach 1 and we have $\Kappa_j \to \matr{I}_Q$ and hence complete shrinkage. When $\tau \to \infty$ all eigenvalues of $\Kappa_j$ approach 0 and we have $\Kappa_j \to \matr{0}_Q$, where $\matr{0}_Q$ denotes a matrix of zeros, and hence no shrinkage.

For an unstructured error variance matrix $\matr{\Sigma}$, a closed form solution for the eigenvalues of the shrinkage factor matrix $\Kappa_j$ is not available. However, it can be instructive to consider simpler parametric forms. For the vector autoregressive model in Section~\ref{sec:model}, we assume that $\matr{\Sigma}^{-1}$ is a symmetric, circulant, tridiagonal matrix, taking the form~\eqref{eq:tri_diag_matrix}. In this case, letting $d_j^2 = N s_j^2\tau^2\sigma_0^{-2}\omega$ and $\tilde{\lambda}_{jk} = d_j \lambda_{jk}$ for $k=1,\dots,K$ we have
\begin{equation*}
\Kappa_j^{-1} =
\begin{pmatrix}
1 + \omega^{-1} \tilde{\lambda}_{j1}^2  &\sigma_0^{2} \tilde{\lambda}_{j1}^2  &0        &0        &\cdots &0      &0        &0        & \sigma_0^{2} \tilde{\lambda}_{j1}^2 \\
   \sigma_0^{2} \tilde{\lambda}_{j2}^2  &1 +    \omega^{-1} \tilde{\lambda}_{j2}^2  &   \sigma_0^{2} \tilde{\lambda}_{j2}^2  &0        &\cdots &0      &0        &0        &0\\
0        &   \sigma_0^{2} \tilde{\lambda}_{j3}^2  &1 +    \omega^{-1} \tilde{\lambda}_{j3}^2  &   \sigma_0^{2} \tilde{\lambda}_{j3}^2  &\cdots &0      &0        &0        &0\\
\vdots   &\vdots   &\vdots   &\vdots   &\ddots &\vdots &\vdots   &\vdots   &\vdots\\
0        &0        &0        &0        &\cdots &0      &   \sigma_0^{2} \tilde{\lambda}_{j,K-1}^2  &1 +    \omega^{-1} \tilde{\lambda}_{j,K-1}^2  &   \sigma_0^{2} \tilde{\lambda}_{j,K-1}^2 \\
   \sigma_0^{2} \tilde{\lambda}_{jK}^2  &0        &0        &0        &\cdots &0      &0        &   \sigma_0^{2} \tilde{\lambda}_{jK}^2  &1 +    \omega^{-1} \tilde{\lambda}_{jK}^2 
\end{pmatrix}.
\end{equation*}
Although a closed-form solution for its eigenvalues and inverse are not available, in the special case when $\omega=0$, so that $\matr{\Sigma}=\sigma_0^{-2} \matr{I}_Q$, the shrinkage factor matrix reduces to
\begin{equation}\label{eq:K_diag}
\Kappa_j = \diag\left( \frac{1}{1 + N s_j^2 \tau^2 \lambda_{j1}^2 \sigma_0^{-2}}, \ldots, \frac{1}{1 + N s_j^2 \tau^2 \lambda_{jQ}^2 \sigma_0^{-2}}\right), \quad j = 1,\ldots,P
\end{equation}
so that each component of the posterior mean can be expressed as a product of a single shrinkage factor and the corresponding element of the least squares estimator
\begin{equation*}
m_{jk} = \left( 1 - \frac{1}{1 + N s_j^2 \tau^2 \lambda_{jk}^2 \sigma_0^{-2}} \right) \hat{a}_{jk}, \quad j=1,\ldots,P, \; k=1,\ldots,Q.
\end{equation*}
Clearly the eigenvalues of $\Kappa_j$ are simply its diagonal entries with a value near 0 or 1 indicating no shrinkage or complete shrinkage of the corresponding least squares estimate. The results for some other parametric forms are described in Section~S5.2 the Supplementary Materials.

\subsection{Effective number of non-zero coefficients}\label{ssec:meff}

In the case of a univariate response, there is a single shrinkage factor $\kappa_j$ for each regression coefficient $a_j$. In the joint prior induced by the half-Cauchy densities for the local shrinkage parameters, each $\kappa_j$ is conditionally independent of $\kappa_k$ ($k \ne j$) given the global shrinkage parameter $\tau$ and the error variance $\sigma^2$. The conditional priors for the $\kappa_j$ can be derived in closed form and have $u$-shaped densities over the unit interval meaning that, \textit{a priori}, most $\kappa_j$ are either zero or one and so
\begin{equation*}
m_{\text{eff}} = \sum_{j=1}^P (1 - \kappa_j)
\end{equation*}
can be interpreted as the effective number of non-zero coefficients.

In the multivariate case, when the error variance $\matr{\Sigma}$ is diagonal, Section~\ref{subsec:shrinkage_factor} showed that the shrinkage factor matrices $\Kappa_j$ are also diagonal. It follows by direct analogy with the univariate result that the priors for the diagonal elements of $\Kappa_j$ will be independent \textit{a priori} with $u$-shaped densities. Therefore the same logic applies and we can interpret
\begin{equation}\label{eq:m_eff}
m_{\text{eff}} = \sum_{j=1}^P \tr\left(\matr{I}_Q - \Kappa_j\right)
\end{equation}
as the effective number of non-zero coefficients. In the case of an unstructured error variance matrix $\matr{\Sigma}$,  conditional on $\matr{\Sigma}$ and $\tau$, the mapping from $Q$-dimensional $(\lambda_{j1}, \ldots, \lambda_{jQ})$ to $Q^2$-dimensional $\Kappa_j$ is dimension-increasing, and so $\Kappa_j$ must lie on a $Q$-dimensional manifold of $\mathbb{R}^{Q^2}$. As such, the independent, unit-median half-Cauchy distributions for the diagonal elements of $\matr{\Lambda}_j$ induce a joint distribution for $\Kappa_j$ for which a density function does not exist. We can, nevertheless, explore the marginal and pairwise joint densities of the elements $\Kappa_{j,k\ell}$ by simulation. For a range of values for the correlations in $\matr{\Sigma}$, a simulation experiment described in Section~S5.3 the Supplementary Materials reveals that the prior distribution for the elements of $\Kappa_j$ assigns high probability to diagonal binary matrices. It does not, therefore, seem unreasonable to continue to interpret $m_{\text{eff}}$, as defined in~\eqref{eq:m_eff}, as the effective number of non-zero coefficients.

The prior expectation of the effective number of non-zero coefficients $m_{\text{eff}}$, conditional on $\tau$ and $\matr{\Sigma}$ is not generally available in closed form owing to the absence of a closed form expression for $\Kappa_j$. However, in the special case when $\matr{\Sigma} = \sigma^2 \matr{I}_Q$, it follows immediately from the results in the univariate case that
\begin{equation*}
\E_{\Lambda | \tau, \Sigma}(m_{\text{eff}}) = \frac{\sqrt{N} \tau \sigma^{-1}}{1 + \sqrt{N} \tau \sigma^{-1}} PQ,
\end{equation*}
where we have assumed that the explanatory variables have been standardised to have variance equal to 1, that is, $s_j^2 = 1$ for $j=1,\ldots,P$. Suppose $e_0$ is our prior expectation for the number of non-zero coefficients in $\matr{A}$. Then we can set $\E_{\Lambda | \tau, \Sigma}(m_{\text{eff}}) = e_0$, $\tau=\tau_0$ and solve for $\tau_0$ to obtain
\begin{equation*}
\tau_0 = \frac{e_0}{PQ - e_0} \frac{\sigma}{\sqrt{N}}
\end{equation*}
which demonstrates that $\tau$ should scale with $\sigma / \sqrt{N}$ if the prior expectation of the effective number of non-zero coefficients $m_{\text{eff}}$ is to remain constant. We can then choose as our prior for $\tau$
\begin{equation}\label{eq:prior_tau}
\tau | \sigma \sim \hcauchy(0, \tau_0^2)
\end{equation}
which has conditional median equal to $\tau_0$ given $\sigma$.

Of course, for general problems it may not be the case that we wish to restrict $\matr{\Sigma} = \sigma^2 \matr{I}_Q$; indeed this is not our choice for the vector autoregression in Section~\ref{sec:model}. In such cases, we can make our prior specification consistent with a central value in the prior for $\matr{\Sigma}$ by introducing a hyperparameter $\sigma$ such that, say, $\E(\matr{\Sigma} | \sigma) = \sigma^2 \matr{I}_Q$ or $\E(\matr{\Sigma}^{-1} | \sigma) = 1 / \sigma^2 \matr{I}_Q$. We can then construct our prior for $(\tau, \matr{\Sigma}, \sigma)$ or $(\tau, \matr{\Sigma}^{-1}, \sigma)$ hierarchically so that
\begin{equation*}
\pi(\tau, \matr{\Sigma}, \sigma) = \pi(\tau | \sigma) \pi(\matr{\Sigma} | \sigma) \pi(\sigma)
\end{equation*}
or
\begin{equation}\label{eq:prior_for_tau_Sig_sig}
\pi(\tau, \matr{\Sigma}^{-1}, \sigma) = \pi(\tau | \sigma) \pi(\matr{\Sigma}^{-1} | \sigma) \pi(\sigma)
\end{equation}
with the conditional distribution for $\tau | \sigma$ specified in~\eqref{eq:prior_tau}.

For the symmetric, circulant, tridiagonal precision matrix $\matr{\Sigma}^{-1}$ in~\eqref{eq:tri_diag_matrix} used in the vector autoregressive model, parameterised in terms of $\varpi_0 = \sqrt{2} ( \sigma_0^{-2} + 2 \omega ) / 2 > 0$ and $\varpi_1 = \sqrt{2} ( \sigma_0^{-2} - 2 \omega ) / 2 > 0$, we use a prior of the form~\eqref{eq:prior_for_tau_Sig_sig} and require
\begin{equation*}
\E(\sigma_0^{-2} | \sigma) = \frac{\E(\varpi_0 | \sigma)+\E(\varpi_1 | \sigma)}{\sqrt{2}} = \frac{1}{\sigma^2} \quad \text{and} \quad \E(\omega | \sigma) = \frac{\E(\varpi_0 | \sigma)-\E(\varpi_1 | \sigma)}{2 \sqrt{2}} = 0
\end{equation*}
and hence
\begin{equation*}
\E(\varpi_0 | \sigma) = \E(\varpi_1 | \sigma) = \frac{\sqrt{2}}{2 \sigma^2}.
\end{equation*}
Specifically, for $i=0,1$, independently, we take
\begin{equation}\label{eq:prior_varpi}
\varpi_i | \sigma \sim \gam\left(\frac{1}{c_{\varpi}^2}, \frac{\sqrt{2} \sigma^2}{c_{\varpi}^2}\right)
\end{equation}
 and then
\begin{equation}\label{eq:prior_sig}
\sigma \sim \lognorm(m_{\sigma}, s_{\sigma}^2).
\end{equation}
In the application in Section~\ref{sec:app}, we choose $c_\varpi = 1, m_\sigma=0$ and $s_\sigma = \sqrt{10}$. When $\matr{\Sigma}$ is unstructured, suitable choices for its prior are outlined in Section~S6 of the Supplementary Materials.

\subsection{Regularised horseshoe prior}
The (original) horseshoe prior in~\eqref{eq:horseshoe} can pose some problems for the inferential scheme if one or more coefficient $a_{jk}$ is only weakly identified by the likelihood. In such cases, the prior imparts little influence and the posterior remains long-tailed and difficult to sample. In order to address this problem, without compromising the interpretation of the effective number of non-zero coefficients, \citet{piironen} propose a modification to the horseshoe, called the regularised horseshoe, which would involve replacing~\eqref{eq:horseshoe} with
\begin{equation} \label{eq:rhs_prior_reg}
a_{jk} | \lambda_{jk}, \tau, c \sim \norm\left(0, \tau^2 \tilde{\lambda}_{jk}^2 \right), \qquad \tilde{\lambda}_{jk}^2 = \frac{c^2\lambda_{jk}^2}{c^2+\tau^2\lambda_{jk}^2}, \qquad
\lambda_{jk} \sim \hcauchy\left( 0, 1 \right), \qquad
\tau \sim \hcauchy\left( 0, \tau_0^2 \right),
\end{equation}
for $j=1,\ldots,P, \; k=1,\ldots,Q$. This can be regarded as a continuous analogue to the replacement of a spike-and-slab prior with infinite slab variance with one with finite variance $c^2$. The parameter $c > 0$ can be fixed or given a prior to reflect beliefs about the maximum possible size of the regression coefficients. We use precisely a prior of this form for the parameters in the autoregressive coefficient matrix $\matr{A}$ in our VAR(1) model, choosing $e_0=\E_{\Lambda | \tau, \Sigma}(m_{\text{eff}}) = 12$, to reflect our beliefs that only the $K=12$ diagonal elements of $\matr{A}$ are likely to be non-zero. We then choose
\begin{equation}\label{eq:prior_c2}
c^2 \sim \invgam(2, 8)
\end{equation}
to reflect the idea that, after scaling the data so that each bin has standard deviation roughly equal to 1, we would not expect any regression coefficient to exceed around 5 in absolute value.  Note that $\invgam(g,h)$ denotes an inverse gamma distribution with shape $g$ and scale $h$.

\section{Bayesian inference}

\subsection{Prior distribution}

There are several parameters in the hierarchical VAR(1) model.  First, there is the matrix of autoregressive coefficients $\matr{A}$ with its global shrinkage parameter $\tau \in \mathbb{R}_+$, the matrix of local shrinkage parameters $\matr{\Lambda}^*$ and the regularising parameter $c^2$.  Additionally, there are the parameters for the precision matrix $\matr{\Sigma}^{-1}$ of the errors, denoted by $\vec{\varpi} = \{\varpi_0, \varpi_1\} \in \mathbb{R}^2_+$ and the parameter $\sigma$ in their hierarchical prior. There are the harmonic regression coefficients for $\vec{\mu}_t$, denoted by $\vec{\theta} = \{ \vec{\beta}_1,\vec{\beta}_2, \vec{\gamma}_1, \vec{\gamma}_2\}$.  We also have the coefficients of the CE covariates $\matr{X}$ and the intercepts in $\matr{B}$, with unknown means $\vec{\mu}_B=(\mu_{B_0}, \ldots, \mu_{B_L})$ and variances $\vec{\sigma}_B=(\sigma^2_{B_0}, \ldots, \sigma^2_{B_L})$ for the $L+1$ rows of $\matr{B}$.  Finally, we have the parameters in $\matr{\Phi}_X=\diag(\phi_{X,1}, \ldots, \phi_{X,5})$ and $\matr{\Sigma}_X$ in the missing data model.

We adopt a prior distribution in which these various parameter blocks are independent, with hierarchical structure within blocks:
\begin{align}
\pi \left(\matr{A},\tau, \matr{\Lambda}^*, c^2, \vec{\varpi}, \sigma, \vec{\theta}, \matr{B}, \vec{\mu}_B, \vec{\sigma}_B,\matr{\Phi}_X, \matr{\Sigma}_{X} \right) &=  \pi\left( \matr{A}|\matr{\Lambda}^*,\tau,c^2 \right) \pi\left(\tau|\sigma\right) \pi(\sigma) \pi(\matr{\Lambda}^*) \pi(c^2) \pi \left(\vec{\varpi}|\sigma\right) \notag\\
&\qquad \times \pi\left(\vec{\theta}\right) \pi\left(\matr{B}|\vec{\mu}_B, \vec{\sigma}_B \right) \pi\left( \vec{\mu}_B \right) \pi\left( \vec{\sigma}_B \right)\label{eq:overall_prior} \\
&\qquad \qquad \times \pi\left( \matr{\Phi}_X \right)\pi \left( \matr{\Sigma}_{X} \right).\notag
\end{align}
Our main focus in this paper has been the joint distribution encoded through the right-hand-side of the first line of~\eqref{eq:overall_prior}. The distributions $\pi\left( \matr{A}|\matr{\Lambda}^*,\tau,c^2 \right)$, $\pi\left(\tau|\sigma\right)$ and $\pi(\matr{\Lambda}^*)$ are given in \eqref{eq:rhs_prior_reg}; $\pi(c^2)$ in~\eqref{eq:prior_c2}; $\pi \left(\vec{\varpi}|\sigma\right)=\pi(\varpi_0 | \sigma) \pi(\varpi_1 | \sigma)$ and $\pi(\sigma)$ in~\eqref{eq:prior_varpi} and~\eqref{eq:prior_sig}, respectively. 

For the parameters in the harmonic regression component of the time-varying mean, we take $\vec{\beta}_j \sim \norm_K(\vec{0},\matr{V}_\beta)$ and $\vec{\gamma}_j \sim \norm_K(\vec{0},\matr{V}_\gamma)$ independently for $j=1,2$, with $\matr{V}_\beta = \matr{V}_\gamma = 100 \matr{I}_K$. For the coefficients of the CE covariates, we adopt hierarchical priors for the $b_{\ell k}$, such that for each $\ell=0,\ldots,L$ independently, we have $b_{\ell k} | \mu_{B_\ell}, \sigma^2_{B_\ell} \sim \norm\left(\mu_{B_\ell}, \sigma^2_{B_\ell}\right)$ independently for $k=1,\ldots,K$, $\mu_{B_\ell} \sim \norm\left(a_\alpha,b_\alpha^2\right)$ and $\sigma^2_{B_\ell} \sim \invgam \left( c_\alpha, d_\alpha \right)$.  To give $\E(b_{\ell k}) = 0$, $\Corr \left( b_{\ell j}, b_{\ell k} \right) = 0.95$ and $\Var \left( b_{\ell j} \right) = 100$, we select $a_\alpha = 0$, $b_\alpha = \sqrt{95}$, $c_\alpha = 2.25$ and $d_\alpha = 6.25$.

Finally, for the parameters in the missing data model, we adopt the prior $\phi_{X,\ell} \sim \bet \left( a_\phi, b_\phi \right)$ independently for $\ell=1,\ldots,5$ and $\matr{\Sigma}_{X} \sim \mathcal{W}^{-1} \left( \matr{S}_{X} , v_{X} \right)$, with $a_\phi=b_\phi=2$,  $\matr{S}_{X}=\matr{I}_5$ and $v_{X}=9$. Note that $\mathcal{W}^{-1}\left(\matr{\Psi},\nu\right)$ denotes an inverse-Wishart distribution with scale matrix $\matr{\Psi}$ and $\nu$ degrees of freedom.

\subsection{Posterior distribution and computation}

Combining the prior~\eqref{eq:overall_prior} with the first-order Markovian likelihood~\eqref{eq:VAR(1)_with_mean} and the missing data model~\eqref{eq:missingdata_model} using Bayes theorem yields the posterior distribution as
\begin{align*}
\pi \left(\matr{A}, \tau, \matr{\Lambda}^*, c^2, \vec{\varpi}, \sigma, \vec{\theta}, \matr{B}, \vec{\mu}_B, \vec{\sigma}_{B}, \matr{\Phi}_X, \matr{\Sigma}_{X} | \matr{Y}, \matr{X} \right) &\propto \pi \left(\matr{A},\tau, \matr{\Lambda}^*, c^2, \vec{\varpi}, \sigma, \vec{\theta}, \matr{B}, \vec{\mu}_B, \vec{\sigma}_{B}, \matr{\Phi}_X, \matr{\Sigma}_{X}\right) \\
&\qquad \times \pi\left(\matr{Y}| \matr{A}, \vec{\theta},\matr{B}, \vec{\varpi}, \matr{X} \right) \pi\left(\matr{X}| \matr{\Phi}_X, \matr{\Sigma}_{X} \right),
\end{align*}
where the likelihood can be written as
\begin{equation*}
\pi\left(\matr{Y}| \matr{A}, \vec{\theta}, \matr{B}, \vec{\varpi}, \matr{X} \right) = \prod_{t=2}^N \pi \left( \vec{y}_t | \vec{y}_{t-1}, \matr{A}, \vec{\theta}, \matr{B}, \vec{\varpi}, \vec{x}_{t-1} \right)
\end{equation*}
with $\vec{y}_t | \vec{y}_{t-1}, \matr{A}, \vec{\theta}, \matr{B}, \vec{\varpi}, \vec{x}_{t-1} \sim \norm_{K}\left( \vec{\mu}_{t} + \matr{A} \left( \vec{y}_{t-1}-\vec{\mu}_{t-1} \right),\matr{\Sigma}\right)$. Here we have implicitly conditioned on the observed value of $\vec{y}_1$ so that the contribution of the marginal model $\pi\left(\vec{y}_1 | \matr{A}, \vec{\theta}, \matr{B}, \vec{\varpi}, \vec{x}_0 \right)$ can be ignored. This is reasonable as we have a large enough sample size that little information will be lost in doing so. The missing data model $\pi(\matr{X}| \matr{\Phi}_X, \matr{\Sigma}_{X} )$ is constructed in an analogous fashion.

This posterior distribution is analytically intractable, so we resort to sampling from it using Markov chain Monte Carlo methods. More specifically, we use Hamiltonian Monte Carlo (HMC) \citep{Nea11,GC11} which is a gradient-based, auxiliary variable method which is well-suited for inference in hierarchical models \citep[][]{betancourt2015}. The HMC algorithm was implemented using \texttt{rstan} \citep[][]{Sta20}, the R interface to the Stan software \citep[][]{CGH17}. Stan requires users to write a program in the probabilistic Stan modelling language, the role of which is to provide instructions for computing the logarithm of the kernel of the posterior density function. The Stan software then automatically tunes and runs a Markov chain simulation to sample from the resulting posterior. The Stan program used in our application can be found in Section~7 of the Supplementary Materials.

We ran the algorithm for 10K iterations, with a warm-up period of 3000 iterations.  In the interests of saving memory, the output is thinned to leave us with 1000 samples from the posterior.  The usual graphical and numerical diagnostic checks  gave no evidence of any lack of convergence and mixing was good. We present the results in the next section based on these posterior samples. 
\section{Application to WWTP data}\label{sec:app}
Now we discuss our findings after fitting our model to the data.  We look at the posterior means and $95\%$ credible intervals (CIs) of the model parameters.  If zero is contained in a CI, we use this as a discriminator to suggest that the parameter's value may be (close to) zero.  However, we emphasise that this does not necessarily mean that there is not considerable support for a positive or negative coefficient.
\subsection{Time varying mean}\label{ssec:AS_mean_results}
We begin with the time varying mean $\vec{\mu}_t$ of the model, which is described by two seasonal harmonics and time-varying CE covariates. Recall that $\vec{\mu}_t = \vec{\alpha}_t + \sum^{2}_{j=1}\vec{\beta}_j \sin\left(2\pi j t/52 \right) + \sum^{2}_{j=1}\vec{\gamma}_j \cos\left( 2\pi j t/52  \right)$, where $\vec{\alpha}_t$ has $k$-th element $\alpha_{tk} = b_{0k} + b_{1k}x_{t-1,1}+\dots+b_{Lk}x_{t-1,L}$ and $\vec{X}_t$ are the transformed measurements of our chosen CE covariates at time $t$.
 
\subsubsection{Chemical and environmental covariates} 
There are five covariates in the model: nitrate, COD, ammonia, pH and phosphate.  For all 12 bins, the CIs for nitrate and phosphate shown in Supplementary Figure~S5 all include zero, suggesting that neither of these covariates has a linear relationship with the time varying mean of any of the bins, although for phosphate we note that the CIs for bins 2 and 3 only just overlap zero. Despite all of the CIs overlapping zero, there is a clear pattern for phosphate, with ``winter blooming'' bins showing a positive relationship and ``summer blooming'' bins showing a negative relationship, though this might just be an artefact of how the data were binned. 

\begin{figure*}[h]
\makebox[\textwidth][c]{\includegraphics[width = 0.66\textwidth]{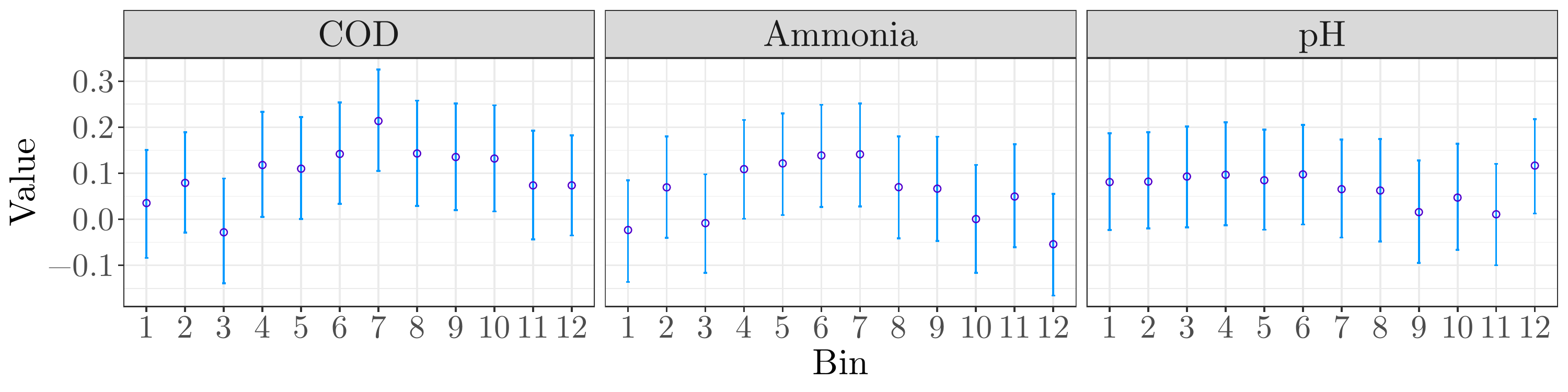}}
\caption{Posterior means ($\color{mypurple} \scriptstyle{\circ}$) and 95\% credible intervals (\textcolor{myblue}{---}) for COD, ammonia and pH.}
\label{fig:AS_chems_in_error_bars.pdf}
\end{figure*}

From Figure~\ref{fig:AS_chems_in_error_bars.pdf}, bins 4 to 10 have positive regression coefficients with COD, with bin 7 having the largest regression coefficient with a posterior mean of 0.2136 (4 d.p.).  Ammonia has a positive regression coefficient with four bins (4 to 7) and, as we saw with COD, bin 7 has the largest regression coefficient with a posterior mean of 0.1412 (4 d.p.).  Finally, bin 12 seems to have a positive relationship with pH, with its regression coefficient having a posterior mean of 0.1167 (4 d.p.).

Removal of ammonia and other pollutants is essential in the treatment of wastewater and ammonia is removed through nitrification by bacteria. In the nitrogen cycle, nitrification is a two-step process of ammonia oxidation then nitrite oxidation.  Bacteria from the genus \textit{Nitrosomonas} can oxidise ammonia to nitrite \citep{wetzel2001}, although there are other ammonia oxidising microorganisms (AOM) too.  \textit{Nitrobacter} bacteria from the same phylum as \textit{Nitrosomonas} oxidise nitrite to nitrate but are difficult to detect \textit{in-situ}. \citet{wagner1996} suggested that this could be because they have a minor role in WWTPs and although \citet{alawi2009} agreed that their role is small, they also noted that lack of detection does not necessarily mean lack of presence. In the AS, no \textit{Nitrobacter} counts are recorded.  This could suggest that in our WWTP other nitrite oxidising bacteria (NOB), for example, \textit{Nitrospira} and \textit{Ca. Nitrotoga}, are responsible for nitrite removal or that the \textit{Nitrobacter} bacteria simply have not been detected, as seen in the literature.

Until recently, \textit{Nitrospira} were considered solely NOB \citep{mehrani2020}. \citet{daims2015} and \citet{vanKessel2015} independently discovered a single microorganism from the genus \textit{Nitrospira} that can carry out complete nitrification through the \textit{comammox} (complete oxidation of ammonia to nitrate) process. Additionally, it has been found that there is a reciprocal feeding interaction between nitrifiers.  Some species of \textit{Nitrospira} are able to convert urea to ammonia and carbon dioxide, which means they can supply AOM with ammonia and in return receive nitrite produced by ammonia oxidation \citep{koch2015}. An OTU from the genus \textit{Nitrospira} is one of the most abundant OTUs within bin 4 and this OTU could be capable of comammox which could provide a reasonable explanation as to why there is a positive coefficient for ammonia and bin 4.   Furthermore, most microorganisms need ammonia to grow via nitrogen assimilation, so that might explain why we see a positive relationship between bins 4 to 7 and ammonia.

To understand why some of the other bins may have a relationship with  COD, we look at the most abundant OTUs within some of the bins.  Supplementary Table~S5 shows the genera of the top six OTUs in each bin.   An OTU from the genus \textit{Terrimonas} is the most abundant in bin 4, with a median within-bin relative abundance of around $22.8\%$. Bacteria from this genus assimilate organic compounds such as sugars and proteins \citep{MiDAS2015}.  This provides a possible explanation as to why bin 4 has a positive relationship with COD. 

Of the 1274 OTUs in bin 5, the most abundant OTU based on median within-bin relative abundance ($\sim 9.1\%$) is from the genus \textit{Zoogloea}.  Bacteria from this genus are highly active oxidisers of organic compounds \citep{dugan1981}.  Recalling that covariates are incorporated into the model via lag-one regression, the transformed COD measurement from the previous time point is used to model the intercept of the time varying mean at the current time point. If COD is high then this would suggest that there is a larger amount of organic compounds available for the \textit{Zoogloea} bacteria to oxidise for energy and grow, thus explaining the positive coefficient between the bin containing \textit{Zoogloea} and COD.  However, this could result in the amount of organic compounds (and COD) decreasing which in turn could eventually slow the growth rate of the \textit{Zoogloea} bacteria. More organic compounds can migrate into the system as more wastewater enters the WWTP which could then cause the COD to rise again.  This describes a predator-prey-like dynamic and demonstrates that the relationships between the covariates and bins (of OTUs) are unlikely to be simple. 

An OTU from the genus \textit{Leptothrix} is the most abundant OTU in bin 7 based on median within-bin relative abundance ($\sim10.4\%$).  Species from this genus typically oxidise iron and manganese \citep{MiDAS2015}. The second most abundant OTU is from the genus \textit{Dechloromonas} with a median within-bin relative abundance of around 10\%. In Section~\ref{ssec:EA_link}, we saw that some of the top genera were correlated with COD (Figure~\ref{fig:heatmap_genera}), where \textit{Dechloromonas} had a fairly weak positive correlation and \textit{Leptothrix} did not appear in the top 12 genera in the AS tank. Some species of \textit{Dechloromonas} are \textit{polyphosphate-accumulating organisms} (PAOs) and some species have a role in \textit{denitrification} \citep{MiDAS2015}.  PAOs are bacteria that aid the removal of organic compounds containing phosphorus from wastewater. 
Denitrification is the reduction of nitrate to the eventual product of nitrogen gas, following a series of intermediate gaseous nitrogen oxide products.  Nitrate and phosphorus both contribute to the COD of wastewater. Applying logic similar to that discussed for the \textit{Zoogloea} bacteria in bin 5, a positive and likely non-linear relationship between COD and bin 7 seems sensible.

Finally, we focus on bin 12, which is the only bin that has a non-zero (positive) coefficient with pH. OTUs 15, 33 and 65, from the genus \textit{Rhodobacter}, represent about $23.9\%$ of bin 12 on average. 
Most \textit{Rhodobacter} strains grow at an optimal pH range of 6.5 - 7.5 \citep{imhoff2015}. The pH ranges from 5.02 to 7.5 with a median of 6.53, thus providing a possible explanation as to why bin 12 has a positive relationship with pH. Figure~\ref{fig:heatmap_genera} in Section~\ref{ssec:EA_link} does not seem to indicate a correlation between \textit{Rhodobacter} and pH.  However, OTU 15 possibly has a weak positive correlation with pH (Figure~\ref{fig:heatmap_OTUs}).  
It is also important to remember that the heatmaps show correlations, not lag-one correlations. Calculating both the correlation (0.1155) and lag-one correlation (0.1679) between pH and \textit{Rhodobacter}, we see that the lag-one correlation is stronger, thus corroborating our results. Furthermore, this relationship remains after allowing for other covariates and interactions, which highlights the benefit of the model; this relationship may otherwise go unnoticed.

\subsubsection{Harmonic regression coefficients}
Now we look at the harmonic regression coefficients of the model. Supplementary Figures S6 and S7 show the posterior means and 95\% CIs for the harmonic regression coefficients $\vec{\beta}_j$ and $\vec{\gamma}_j$ for $j = 1,2$.  The change in the values of the $\beta_{jk}$ and $\gamma_{jk}$ across bins, $k=1,\hdots,12$, for the first harmonic ($j=1$) can be explained by our chosen clustering method, which is based on the idea that the OTUs display seasonal variation and peak in different months.  Recalling from Section~\ref{ssec:ts_clustering}, OTUs in bin 1 peak in February, OTUs in bin 2 peak in January, OTUs in bin 3 peak in December and so on.  Based on the CIs, it would seem that only bins 2, 3 and 4 seem to have non-zero coefficients for the second harmonic, suggesting that their scaled log counts do not follow a pattern as simple as a sinusoid.  
Recall that in the time series plots of the bins in Figure~\ref{fig:AS_bins} (Section~\ref{ssec:ts_clustering}), we saw that the annual peaks were not as obvious in bins 2 and 3, suggesting a sinusoid may not be such a good descriptor.  Perhaps, this is why we have non-zero coefficients for the second harmonics for these two bins.  Reviewing the time series plots again, we can also see that in bin 4, there seem to be two peaks within 2013, with a smaller peak in the middle of the year and a larger peak around October. This might explain why we have a non-zero coefficient for the second harmonic in bin 4.  Supplementary Figure S8 shows posterior means of the time varying means $\vec{\mu}_t$ and the $95\%$ CIs plotted over the scaled log counts for each bin.  The seasonal patterns of each bin seem to have been captured fairly well.   

\subsection{Matrix of autoregressive coefficients}\label{ssec:AS_A_results}
\begin{figure}[!htb]
\centering
\includegraphics[width=0.42\textwidth]{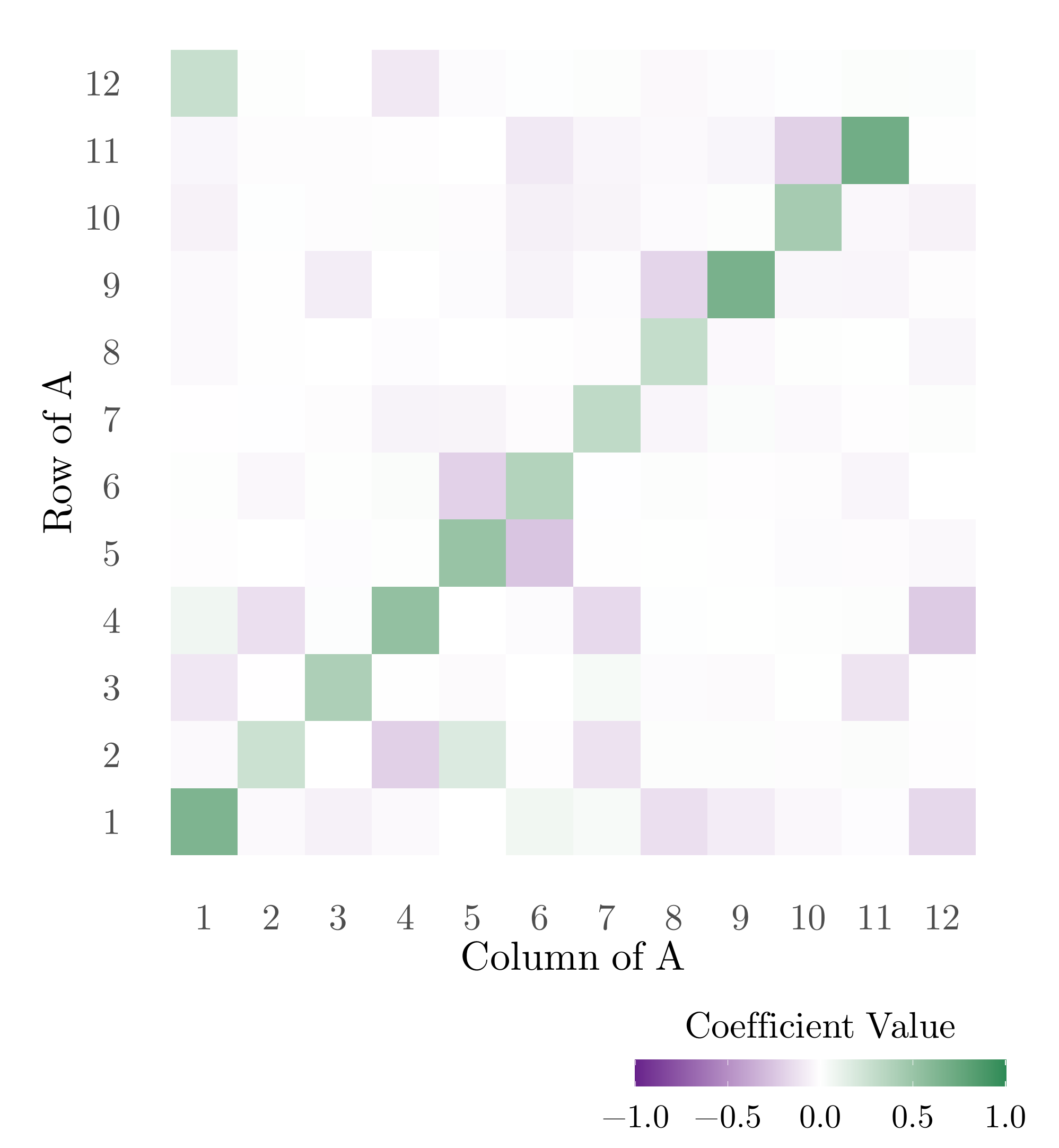}
\caption{Heatmap of the posterior means of the autoregressive coefficients.}
\label{fig:AS_A_heatmap}
\end{figure}
\begin{figure*}[!htb]
\makebox[0.95\textwidth][c]{\includegraphics[width = 0.9\textwidth]{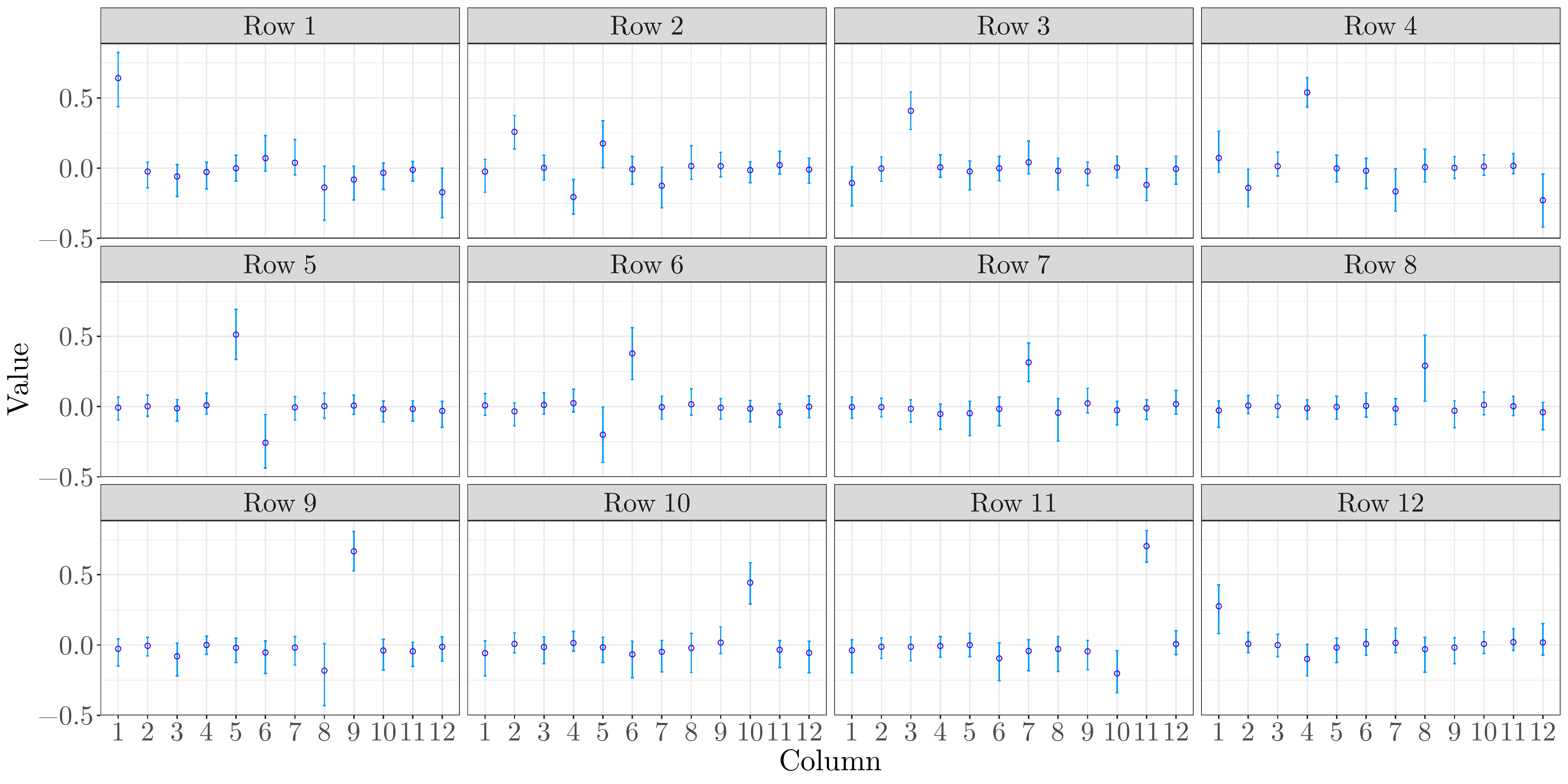}}
\caption{Posterior means ($\color{mypurple} \scriptstyle{\circ}$) and 95\% credible intervals (\textcolor{myblue}{---}) of the autoregressive coefficients.}
\label{fig:AS_A_errorbars}
\end{figure*}

The matrix of autoregressive coefficients is informative about the relationships between bins.  We note that all the posterior samples of $\matr{A}$ have a spectral radius less than one and hence they lie within the stationary region.  This suggests that the system is stable after allowing for seasonality and the effects of the CE covariates. The posterior means of the autoregressive coefficients are shown in a heatmap in Figure \ref{fig:AS_A_heatmap} and they are also shown in Figure~\ref{fig:AS_A_errorbars} with their corresponding 95\% CIs. From the heatmap, we can see that the matrix of autoregressive coefficients based on posterior means is fairly sparse.  With the exception of bin 12, all the bins have a positive autoregressive coefficient with themselves. In other words, the scaled log count of the previous time point seems to have a positive relationship with the scaled log count at the current time point, which seems sensible. Bins 1, 4, 5, 9 and 11 have particularly large ``within-bin'' autoregressive coefficients with posterior means larger than 0.5.  It is surprising that the $a_{12,12}$ is a near-zero coefficient, with a posterior mean of 0.019 (3 d.p.).  It could be that $y_{t,12}$ is better explained by $y_{t-1,1}$ than $y_{t-1,12}$.  Bin 1 peaks in February and bin 12 peaks in March and the posterior mean of $a_{12,1}$ is positive ($0.276$), so this does not seem unreasonable.  The notably larger diagonal values in $\matr{A}$ suggest that a possible improvement to the regularised horseshoe in application to vector autoregressions might allow the diagonal and off-diagonal elements to have their own global shrinkage parameters. 

In addition to the within-bin autoregressive coefficients $a_{kk}$, we see from the CIs in Figure~\ref{fig:AS_A_errorbars} that there is evidence for a few non-zero ``between-bin'' posterior autoregressive coefficients $a_{jk}, j\neq k$. The posterior means for these coefficients whose CIs do not overlap zero are listed in Table~\ref{tab:A_coefs}.  Apart from $a_{12,1}$, the non-zero between-bin coefficients are smaller than all the non-zero within-bin coefficients.

\begin{table}[!htb]
\centering
\begin{tabular}{c c c c} \hline
\textbf{Coefficient}  & \textbf{Posterior Mean} & \textbf{Coefficient}  & \textbf{Posterior Mean}  \\ \hline
$a_{2,4}$  &  -0.204 & $a_{4,12}$  & -0.229 \\
$a_{2,5}$  & \ 0.176 & $a_{5,6}$  & -0.257 \\
$a_{3,11}$  &  -0.119 & $a_{6,5}$  & -0.200 \\
$a_{4,2}$  & -0.14 & $a_{11,10}$  & -0.202 \\
$a_{4,7}$  &-0.166 & $a_{12,1}$  & \ 0.276 \\ \hline 
\end{tabular}
\caption{Posterior means (3 d.p) of the non-zero between-bin coefficients.}
\label{tab:A_coefs}
\end{table}

To aid biological interpretation of the non-zero coefficients we look at the most abundant OTUs in each bin again.  As mentioned above, the most abundant OTU in bin 5 is from the genus \textit{Zoogloea}.  The second most abundant OTU is from the genus \textit{Acidovorax}.  In bin 6, the second most abundant OTU is from the genus \textit{Dechloromonas}, which as mentioned above is capable of nitrite reduction, as well as sulphate reduction.  This is also true for \textit{Zoogloea} and \textit{Acidovorax} bacteria \citep{MiDAS2015}. Perhaps these bacteria amongst others that are not in the most abundant OTUs are competing for resources such as nitrite and sulphate, resulting in the negative autoregressive coefficients.  

As stated above,  an OTU from the strictly aerobic genus \textit{Terrimonas} is the most abundant in bin 4.  The second most abundant OTU is from the genus \textit{Ca. Microthrix}, which is also described as aerobic in \citet{MiDAS2015}. The top two OTUs in bin 2 are from the family \textit{Rhodobacteraceae} with unknown genera. There are at least 288 known species from 99 genera \citep{pujalte2014} in the family \textit{Rhodobacteraceae}, any of which the top two OTUs could be from.  
However, the third most abundant OTU, representing on average $14.5\%$ of bin 2, is from the genus \textit{Haematobacter} from the same family, which are aerobic bacteria.   We cannot determine the genera of the top two OTUs but they may be aerobic, especially as most \textit{Rhodobacteraceae} are aerobic \citep{pujalte2014}.   
Perhaps there are negative interactions between bins 2 and 4 because aerobic microorganisms in both bins are competing for oxygen.   



\subsection{Precision matrix for errors}\label{ssec:AS_prec_results}
Recall that the errors in our model $\vec{\epsilon}_t$ follow a $\norm_K(0,\matr{\Sigma})$ distribution and we have a symmetric, tridiagonal, circulant precision matrix for the errors, shown in \eqref{eq:tri_diag_matrix}. 
The posterior means for $\omega_0$ and $\omega_1$ are 6.7354 and -3.2183 (to 4 d.p.) respectively, with standard deviations 0.1926 and 0.0987 (to 4 d.p.).  
The covariance matrix for the errors $\matr{\Sigma}$ is a symmetric, circulant matrix.  The correlation matrix associated with $\matr{\Sigma}$ is therefore defined by the lag-$k$ correlations $\rho_k$ for $k=1,\hdots,6$.  Supplementary Figure~S9 shows the posterior means and $95\%$ CIs for $\rho_1,\hdots,\rho_6$.  All of the CIs lie above zero which provides evidence of between-bin correlation in the errors.  

\section{Discussion}\label{sec:discuss}
The main aim of this paper was to model the counts of OTUs in the AS of a WWTP and their interactions with each other over time, whilst also allowing for chemical and environmental effects.  Microorganisms in AS are responsible for biologically treating wastewater.  Gaining an understanding of the complex network of microbial interactions is important to ensure a WWTP can continue functioning or, better still, be improved \citep{Cydzik2016}.  As is commonly found in metagenomics studies, our data suffer from high-dimensionality and sparsity.  Owing to the evidence of seasonality in the data, we chose a seasonal phase-based clustering approach to address both issues.

Often, in time series metagenomics, gLV differential equations are used to model non-linear dynamics of the microbial communities of interest.   However, we chose a more parsimonious option, which allows explicit modelling of the error, and developed a Bayesian hierarchical VAR(1) model for our clustered data. This is a simple first-order approximation to a gLV model.  
The circular time-ordering of the bins suggested a sparse autoregressive coefficient matrix would be sensible.  We used a regularised horseshoe prior to allow for this.  The posterior can be very sensitive to the choice of prior for the global shrinkage parameter in the horseshoe prior.  We therefore extended the work of \citet{piironen}, who considered its choice in the context of linear regression for a univariate response, to the multivariate setting.  This gives a principled methodology for constructing the prior based on prior beliefs about the degree of sparsity.  We gave the errors of our model a symmetric, circulant, tri-diagonal precision matrix to complement the chosen clustering method.    To capture the seasonal variation in each bin, we used a harmonic regression to fit a time varying mean, in which the CE data were incorporated. 

After fitting the model to our WWTP data, we identified possible relationships amongst bins  and between bins and CE covariates by inspecting the posterior distributions obtained for the parameters in the model.  For example, we found that some bins seem to have a positive relationship with COD and ammonia.  After looking at the most abundant genera in each bin and biological literature, we also found evidence to suggest that microorganisms may be competing for resources.   Altogether, the analysis provides an interesting insight into the dynamics of the microbial communities present in the AS of the WWTP.
\section*{Acknowledgements}
This work was supported by the Engineering and Physical Sciences Research Council (EPSRC), Centre for Doctoral Training in Cloud Computing for Big Data (grant number EP/L015358/1).  This work was also supported by the EPSRC (grant number EP/N510129/1) via the the Alan Turing Institute project ``Streaming data modelling for real-time monitoring and forecasting''.
\appendix
\section{Supplementary materials}
The supplementary materials related to this article can be found online.
\newpage
  \bibliographystyle{elsarticle-harv}
  \bibliography{paper.bib}

\end{document}